\documentclass[twocolumn,osajnl2,showpacs,amsmath,amssymb,nofootinbib,nobibnotes,floatfix]{revtex4}
\usepackage{graphicx}
\usepackage{dcolumn}
\usepackage{bm}
\usepackage{hyperref}
\usepackage{natbib}
\usepackage{subfigure}
\usepackage{color}

\newcommand{\tr}{\mathrm{tr}}
\begin{document}


\title{Spin squeezing, entanglement and coherence in two driven, dissipative, nonlinear cavities coupled with single and two-photon exchange}
\author{Ali \"U. C. Hardal}
\affiliation{Department of Physics, Ko\c{c} University, \.Istanbul, 34450, Turkey}
\author{\"{O}zg\"{u}r E. M\"{u}stecapl{\i}o\~{g}lu}
\email{omustecap@ku.edu.tr}
\affiliation{Department of Physics, Ko\c{c} University, \.Istanbul, 34450, Turkey}
\begin{abstract}
We investigate spin squeezing, quantum entanglement and second order coherence in two coupled, driven, dissipative, nonlinear cavities. We compare these quantum statistical properties for the cavities coupled with either single or two-photon exchange.
Solving the quantum optical master equation of the system numerically in the steady state, we calculate the zero-time delay second-order correlation function for the coherence, genuine two-mode entanglement parameters, an optimal spin squeezing inequality associated with particle entanglement, concurrence, quantum entropy and logarithmic negativity. We identify regimes of distinct quantum statistical character depending on the relative strength of photon-exchange and nonlinearity. Moreover, we examine the effects of weak and strong drives on these quantum statistical regimes.
\end{abstract}

\maketitle
\section{Introduction}
Multiphoton processes in quantum optical systems have been intensely studied~\cite{hillery2009introduction,dell2006multiphoton} due to their central role in
modern applications such as quantum switching~\cite{ham2000coherence}, quantum communication and computation~\cite{nielsen2010quantum}.
They are also used for fundamental explorations of phase transitions in coupled nonlinear cavity or
superconducting (SC) circuit quantum electrodynamics (QED) systems~\cite{vidal2003entanglement, sondhi1997continuous, PhysRevLett.109.053601,PhysRevA.76.031805,PhysRevLett.99.186401,PhysRevLett.100.216401,
PhysRevA.77.031803,PhysRevLett.103.086403,
PhysRevA.80.023811,PhysRevA.80.033612,PhysRevA.81.061801,PhysRevLett.104.216402,
hartmann2006strongly,greentree2006quantum,PhysRevLett.93.037001}.

An  earlier study of two nonlinear cavities coupled with single-photon exchange~\cite{ferretti2010photon} revealed a curious interplay
between coherence and localization of the photons. It is concluded that photons are coherent and delocalized over the cavities when the tunneling exchange is stronger than the nonlinearity. In the opposite case of weaker tunneling, photons
are localized in each cavity and antibunched~\cite{paul1982photon} .

In the present work, we address the question if such an interplay can go across to quantum correlations as well. We specifically ask how such mutual influences between localization, coherence and quantum entanglement~\cite{horodecki2009quantum} change under two-photon exchange~\cite{alexanian2011two, Hardal12, PhysRevA.85.023833} as well as under strong drive conditions~\cite{PhysRevLett.105.100505,PhysRevA.46.R6801}.
Realization of two-photon exchange coupling was proposed for circuit QED systems~\cite{alexanian2010scattering, PhysRevB.79.180511, PhysRevB.77.180502,niemczyk2011selection,deppe2008two,mariantoni2011photon}.

Profound relations between different definitions of spin squeezing and entanglement have already been explored in detail~\cite{ma2011quantum,marchiolli2013spin,wang_spin_2003,dong_spin_2005,mustecap2011}. Using Lipkin-Meshkov-Glick (LMG) Model~\cite{lipkin1965validity}, interplay of quantum correlations in squeezing and entanglement in finite quantum systems was discussed quite recently~\cite{marchiolli2013spin}.
Entanglement dynamics and exact properties of LMG model in the thermodynamical limit have been carefully analyzed in Refs.~\cite{PhysRevA.70.062304,PhysRevE.78.021106,PhysRevLett.99.050402}.
Two photon exchange coupled cavities can also be described by the LMG model~\cite{Hardal12}. Our present contribution establishes further connections between coherence and mode entanglement to
spin squeezing and particle entanglement in this model under drive and dissipation.

We calculate and compare the zero time delay second order quantum coherence function~\cite{scully1997quantum}, an optimal spin squeezing inequality associated with particle entanglement~\cite{PhysRevLett.99.250405}, l-concurrence~\cite{PhysRevA.64.042315}, genuine mode entanglement parameters~\cite{PhysRevLett.96.050503,PhysRevA.74.032333} and the von Neumann entanglement entropy~\cite{nielsen2010quantum}. Mode entanglement occurs in the second quantization description of the system; and hence it
is fundamentally different from the particle entanglement, which happens in the first quantization description~\cite{cunha2007entanglement,PhysRevA.72.064306,benatti2011entanglement}. Some possible realizations and
experiments to detect multiparticle entanglement via optimal spin squeezing inequalities can be found in Refs.~\cite{prevedel2009experimental,barreiro2010experimental,korbicz2006generalized}.

We have recently examined the same system from the perspective of
dynamical transfer of particle entanglement between photonic and atomic
subsystems~\cite{Hardal12}. We considered ideal evolution of initially entangled and unentangled states
without dissipation and used exact analytical solutions of the model system.
On the other hand, present paper addresses the quantum correlations in the
stationary solutions of the Master equation of the system by taking into
account the open system conditions under drive and dissipation. In addition
to these fundamental and methodological differences between the two
investigations; the objectives of the both work are entirely distinct. Here, the
subtle relations between delocalization and localization and the quantum
coherence are examined from the point of view of quantum correlations.
The inspiring work in Ref.~\cite{ferretti2010photon} reports the localization and coherence relation
in single photon exchange between the cavities. Our results however show
that these relations are strikingly different in the case of two photon
exchange. Ref.~\cite{ferretti2010photon} only considered localization and coherence question but
did not consider if the localization and quantum correlations have similar
interplay; while here we investigate this as well.

Motivation for our purpose is to comprehend the relation between coherence and distinct quantum correlations
and use this relation as a guide to develop viable strategies for practical realization of quantum entanglement in networks of coupled
nonlinear cavities~\cite{hartmann08,liew12}. These networks share characteristic models identical with ultracold atoms in optical 
lattices~\cite{zhou2009pair,PhysRevLett.81.1539,PhysRevLett.81.1543,PhysRevLett.98.030402,ketterle_nature,Pietraszewicz12} and quantum
phase transitions of light~\cite{greentree06} exhibit similar properties to those atomic systems~\cite{jin13}. Novel behaviors could emerge under strong drive
and with strong local and nonlocal nonlinearities in optical systems. Distinct entanglement types under nonlocal nonlinearities and strong drive conditions and their relation
to coherence are not
discussed with sufficient detail in the literature. Filling these gaps could be significant for practical realization of distinct quantum entanglements in photonic cavity networks
for different simulations and applications, as well as for illuminating novel quantum phases in such systems.

This paper is organized as follows. In Sec.~\ref{sec:model} we describe the single and two photon exchange coupled nonlinear cavity models under consideration. In Sec.~\ref{sec:result} we introduce the parameters that we calculate to characterize coherence, entanglement and spin squeezing and present
their steady state results. We conclude in Sec.~\ref{sec:conc}.
\section{The Model System}\label{sec:model}
We consider a system of two identical nonlinear cavities, labeled with $i=1,2$, coupled either by single or two photon exchange interactions. Both cavities are driven by  a coherent pump at rate $F$ at the laser frequency $\omega_{L}$. The corresponding model Hamiltonians  in a frame rotating at $\omega_{L}$ can be written as
\begin{eqnarray}\label{eq:model}
\nonumber\hat{{H}}^{(1)}&=&\sum_{i=1,2}(U\hat{b}_{i}^{\dagger}\hat{b}_{i}^{\dagger}\hat{b}_{i}\hat{b}_{i}+F\hat{b}_{i}^{\dagger}+F^{*}\hat{b}_{i})\\
\label{eq:single}&+&J(\hat{b}_{1}^{\dagger}\hat{b}_{2}+\hat{b}_{2}^{\dagger}\hat{b}_{1}),\\
\nonumber\hat{{H}}^{(2)}&=&\sum_{i=1,2}(U\hat{b}_{i}^{\dagger}\hat{b}_{i}^{\dagger}\hat{b}_{i}\hat{b}_{i}+F\hat{b}_{i}^{\dagger}+F^{*}\hat{b}_{i})\\
\label{eq:two}&+&J(\hat{b}_{1}^{2\dagger}\hat{b}_{2}^{2}+\hat{b}_{2}^{2\dagger}\hat{b}_{1}^{2}),
\end{eqnarray}
where $U$ is the nonlinearity parameter and $J$ is the photon exchange coefficient. Here we denote the model with single and
two photon exchange coupling as $\hat{{H}}^{(1)}$ and $\hat{{H}}^{(2)}$, respectively. These models describe generic two-site Kerr-Hubbard type interactions that may be realized in settings other than coupled cavities, for example
in ultracold atoms~\cite{zhou2009pair,PhysRevLett.81.1539,PhysRevLett.81.1543,PhysRevLett.98.030402,ketterle_nature}.
Two photon exchange term appears in dipolar spinor Bose-Einstein condensates as well~\cite{Pietraszewicz12}. Annihilation (creation) operator for the cavity photons with frequency $\omega_i$ is denoted by
$\hat{b}_i$ ($\hat{b}_i^{\dagger}$). We take $\omega_i=\omega_{L}$. Transformation of the Hamiltonians to the frame rotating at $\omega_L$ removes the explicit time dependence of the drive terms and cancels the $\omega_i b_i^\dag b_i$ terms.

Let us re-express the model Hamiltonians using the pseudo-spin operators of the cavity fields,
\begin{eqnarray}\label{eq:pspin}
\nonumber \hat{J}_{x}&\equiv&\frac{1}{2}(\hat{b}_{1}^{\dagger}\hat{b}_{2}+\hat{b}_{2}^{\dagger}\hat{b}_{1}),\\
 \hat{J}_{y}&\equiv&\frac{-i}{2}(\hat{b}_{1}^{\dagger}\hat{b}_{2}-\hat{b}_{2}^{\dagger}\hat{b}_{1}), \\
\nonumber \hat{J}_{z}&\equiv&\frac{1}{2}(\hat{b}_{1}^{\dagger}\hat{b}_{1}-\hat{b}_{2}^{\dagger}\hat{b}_{2}).
\end{eqnarray}
They satisfy the SU(2) spin algebra $[\hat{J}_\alpha,\hat{J}_\beta]=\epsilon^{\alpha\beta\gamma}\hat{J}_\gamma$. Here $\alpha,\beta,\gamma\in\{x,y,z\}$ and $\epsilon^{\alpha\beta\gamma}$ is the Levi-Civita density. We obtain
\begin{eqnarray}
\label{eq:bjj}\hat{H}^{(1)}&=&U\hat{J}_{z}^2+2J\hat{J}_{x}+\sum_{i=1,2}(F\hat{b}_{i}^{\dagger}+F^{*}\hat{b}_{i}),\\
\label{eq:lmg}\hat{H}^{(2)}&=&U\hat{J}_{z}^2+2J(\hat{J}_{x}^2-\hat{J}_{y}^2)+\sum_{i=1,2}(F\hat{b}_{i}^{\dagger}+F^{*}\hat{b}_{i}).
\end{eqnarray}
While the tunneling term is a mere rotation operator in the case of single photon exchange, it is a generator of spin squeezing that redistributes
the spin fluctuations by twisting them about the two axis~\cite{PhysRevA.47.5138} in the case of two photon exchange. The nonlinear term always acts as a generator of spin squeezing by twisting the fluctuations around a single ($z$) axis. Spin squeezing is associated with multi-particle entanglement. Another type of entanglement can be found between the cavity modes. This mode entanglement is enforced by the mode coupling character of the tunneling terms. Drive term brings coherence into the system.

In the next section we define the measures of quantum correlations to calculate spin squeezing, entanglement and coherence properties of these models. All the measures are closely related to the spin noise and hence strong interplay between coherence, squeezing and types of entanglement is expected.
\begin{figure*}
     \begin{center}
        \subfigure[]{%
            \label{fig:fig1a}
            \includegraphics[width=4cm]{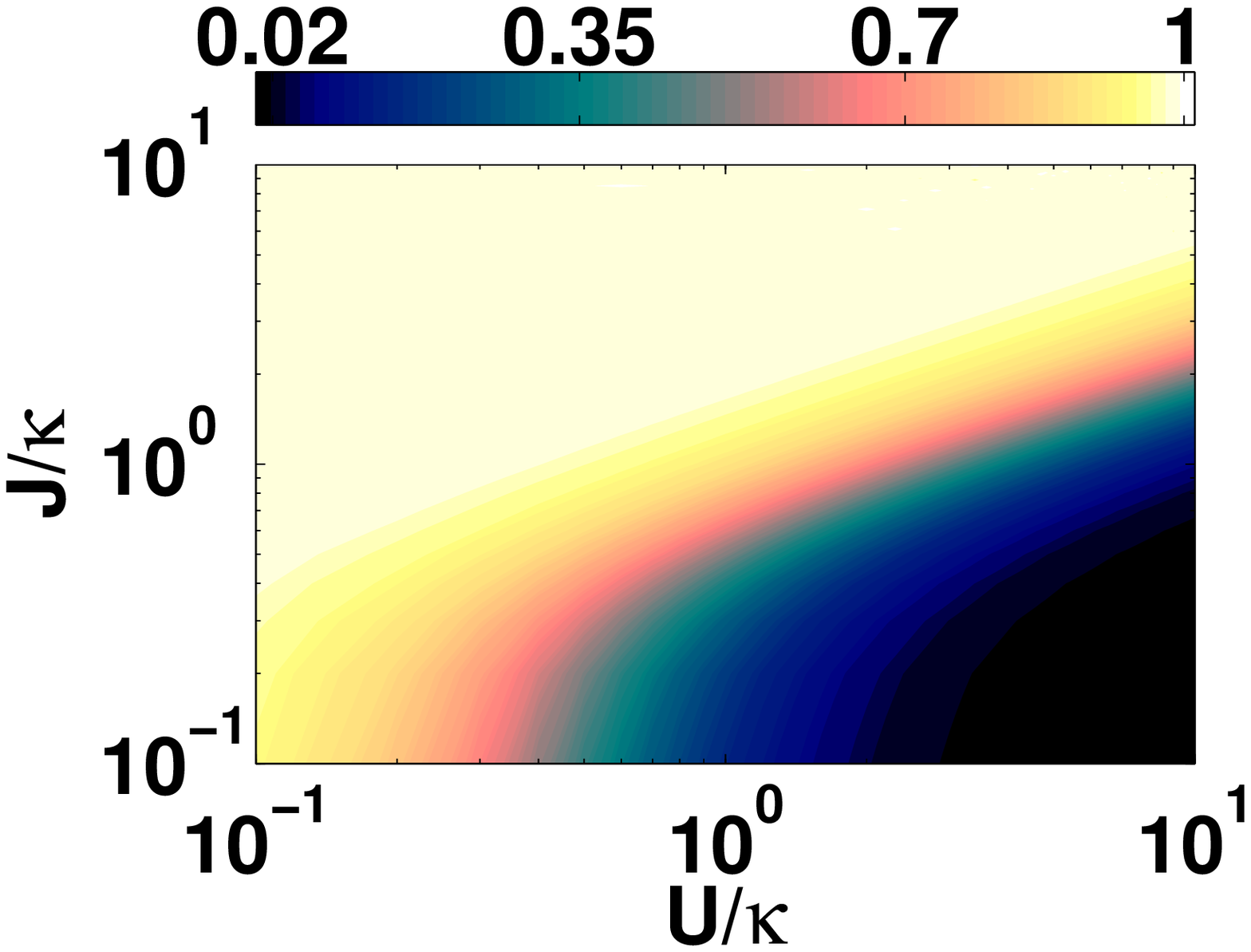}
        }%
        \subfigure[]{%
           \label{fig:fig1b}
           \includegraphics[width=4cm]{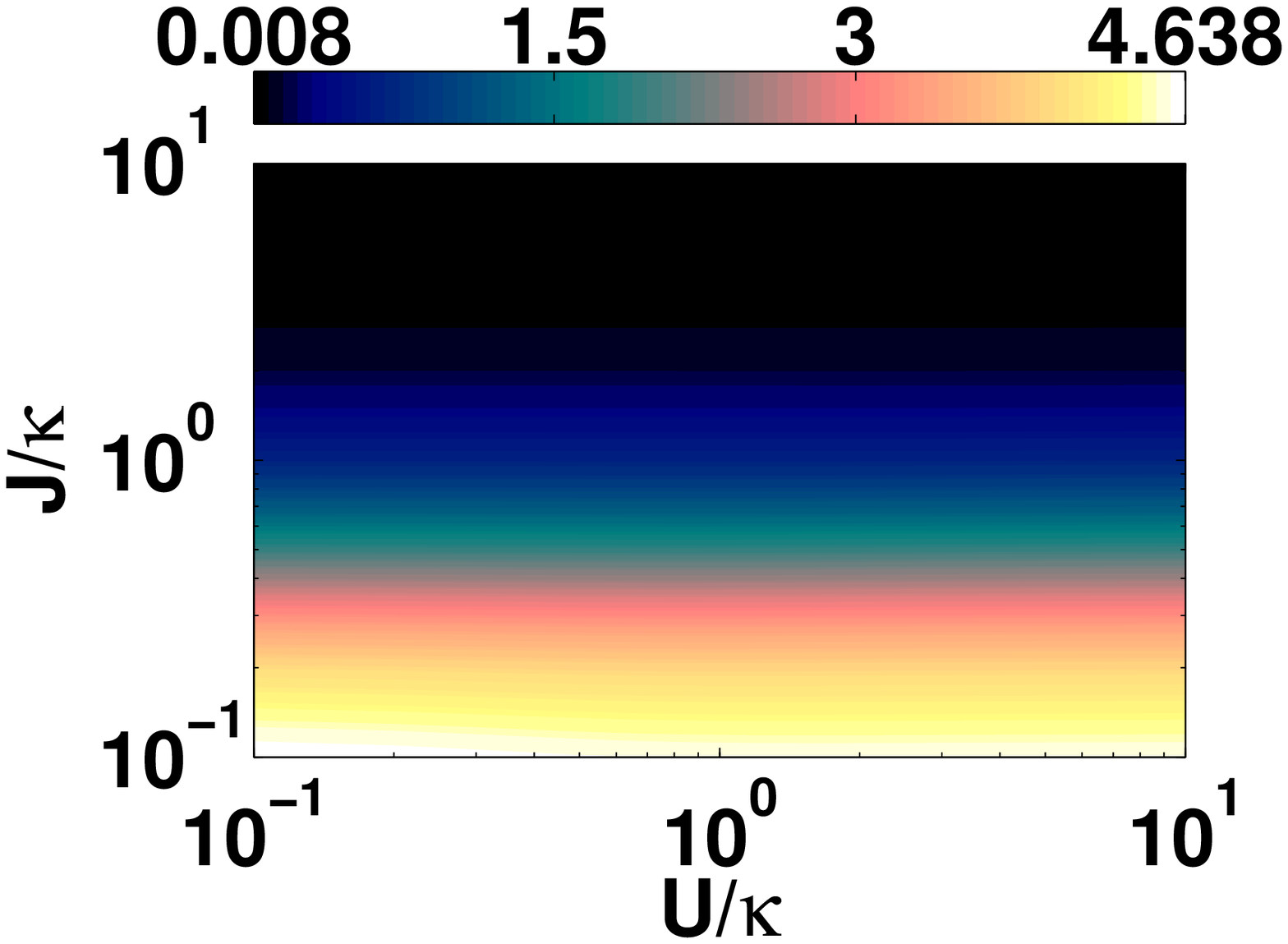}
        }%
           \subfigure[]{%
            \label{fig:fig1c}
            \includegraphics[width=4cm]{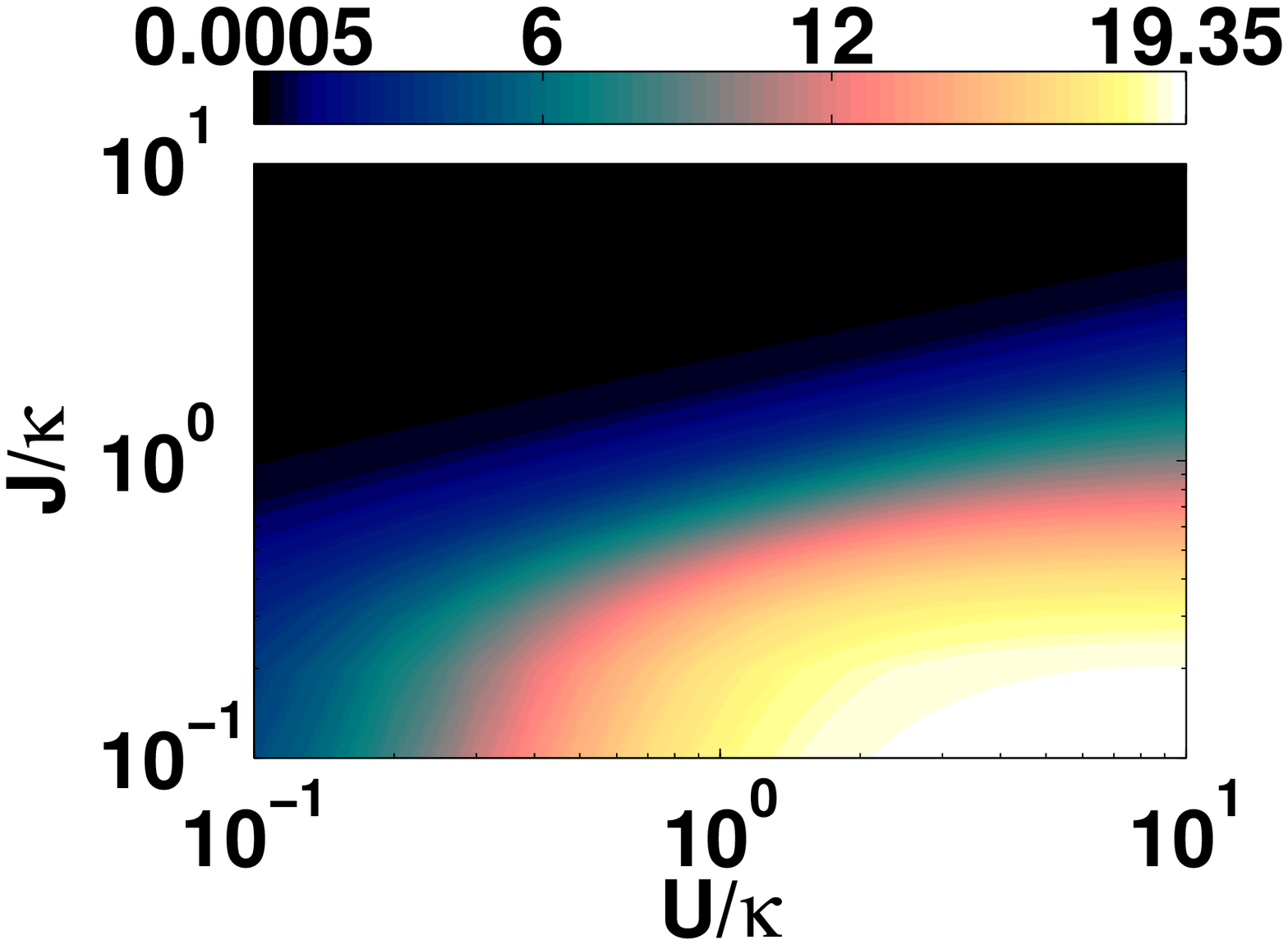}
        }\\ 
        \subfigure[]{%
            \label{fig:fig1d}
            \includegraphics[width=4cm]{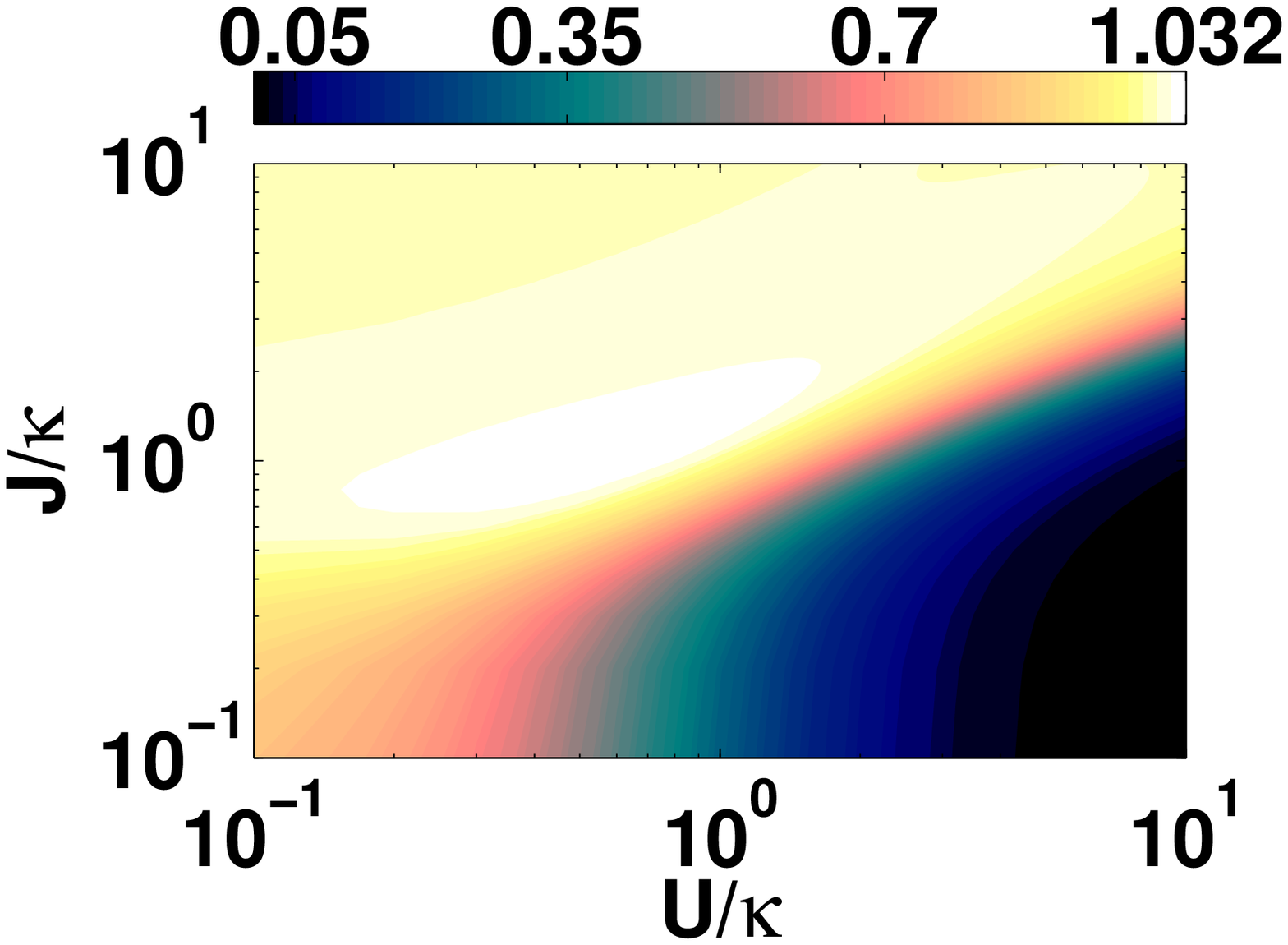}
        }%
        \subfigure[]{%
            \label{fig:fig1e}
            \includegraphics[width=4cm]{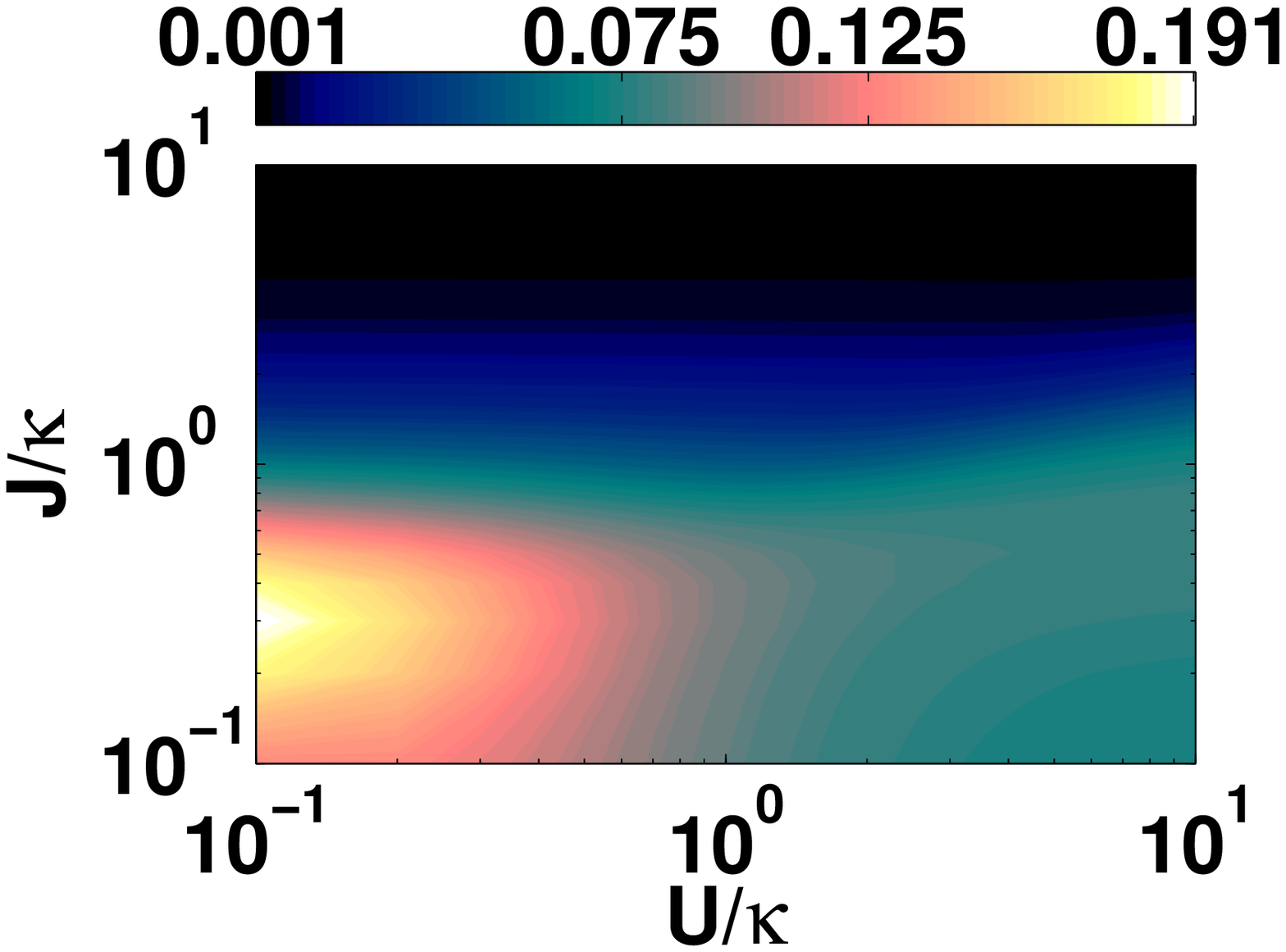}
        }%
        \subfigure[]{%
            \label{fig:fig1f}
            \includegraphics[width=4cm]{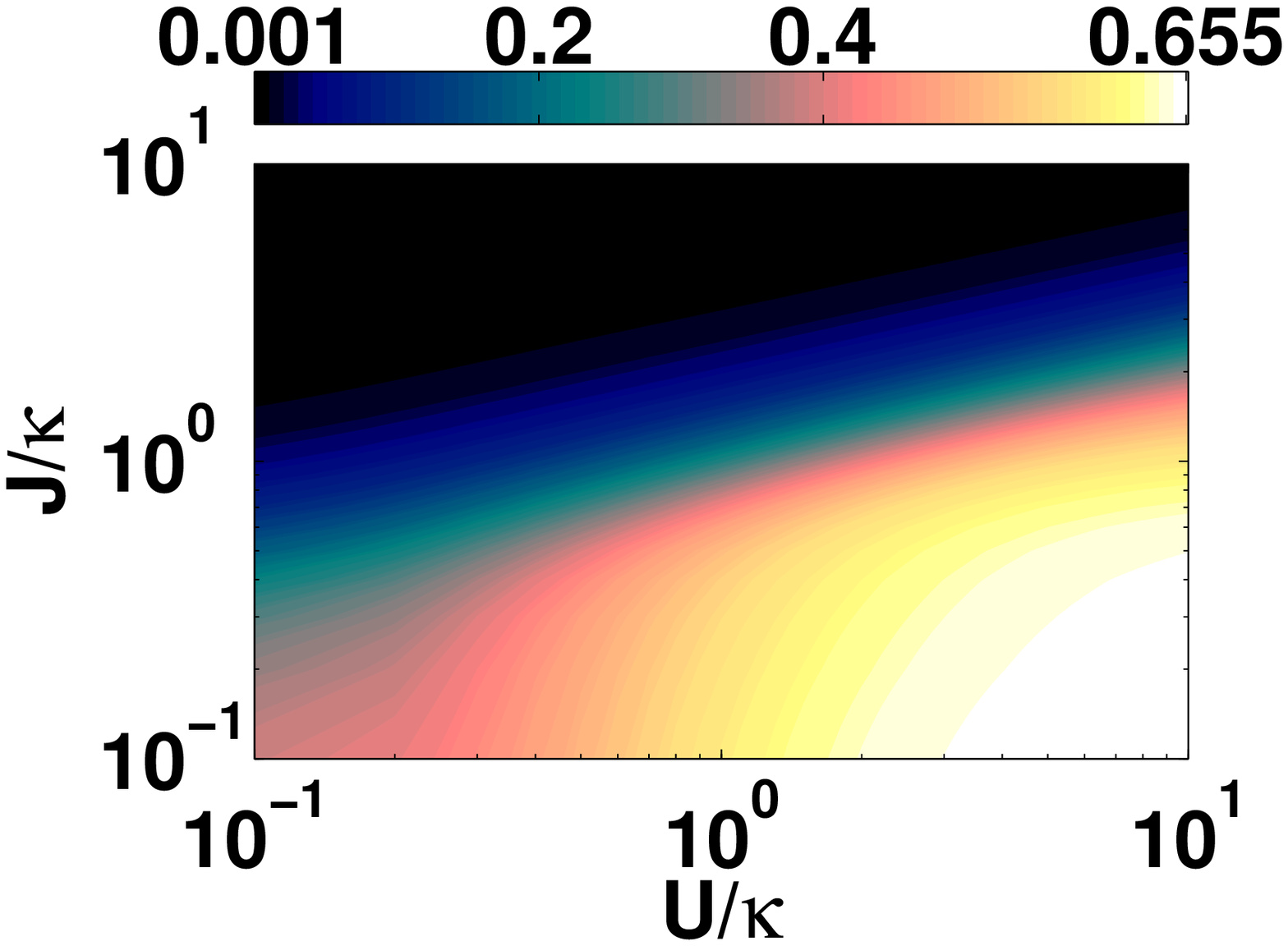}
        }%
    \end{center}
    \caption{%
        (Color Online) Dependence of (a) $g^{(2)}(0)$, (b) $\zeta$ (c) $C_I$ for $F/\kappa=0.1$ and (d) $g^{(2)}(0)$, (e) $\zeta$ (f) $C_{I}$ for $F/\kappa=1$ with respect to dimensionless $J/\kappa$ and $U/\kappa$ in two-site KH system with single-photon exchange. $\zeta$ and $C_{I}$ are multiplied by $10^3$ in the case of $F/\kappa=0.1$ for visibility.
     }%
   \label{fig:fig1}
\end{figure*}
\section{Results and Discussions}\label{sec:result}
To investigate the quantum dynamics of our model systems under dissipation, we assume that the coupling of the cavity photons and the reservoir photons is weak; and the correlation time of the reservoir photons is negligibly short. Under these so called Born and Markov conditions, the dynamics of the open systems can be determined by solving the master equations
\begin{equation}\label{eq:master}
\hat{\dot{\rho}}^{(j)}=-i[\hat{{H}}^{(j)},\hat{\rho}^{(j)}]+\sum_{i=1,2}\kappa_{i}\mathcal{D}[x]\hat{\rho}^{(j)},
\end{equation}
for both single ($j=1$) and two photon ($j=2$) exchange cases. Here,  $\hat{\rho}^{(j)}$
is the density operator for the corresponding case; $\kappa_{i}$ are the photon loss rates out of the cavities. $\mathcal{D}[x]\hat{\rho}^{(j)}=[2\hat{x}\hat{\rho}^{(j)}\hat{x}^{\prime}-\hat{x}^{\prime}\hat{x}\hat{\rho}^{(j)}-
\hat{\rho}^{(j)}\hat{x}^{\prime}\hat{x}]/2$  are the Liouvillian superoperators in the Lindblad form. We assume $\kappa_i\equiv \kappa$ and coherent pump amplitude $F$ is taken to be real. Nonlinearity in our model Hamiltonians is assumed due to a Kerr type nonlinear material in the cavities and thus we can use the bare dissipation rates of the cavities in our treatment. Our generic models could effectively describe other systems such as those of circuit QED in dispersive regime~\cite{PhysRevA.79.013819} or exciton-polariton systems of coupled cavity QED~\cite{ferretti2010photon}. In such cases, canonical transformations used to obtain the effective models apply to the Lindblad noise terms and 
dressed dissipation rates must be used~\cite{ferretti2010photon,PhysRevA.79.013819}.

We solve the master equation for the steady state density operator using the QuTiP: Quantum Toolbox in Python software~\cite{johansson_qutip_2013}. The master equation is solved by taking $N_{1}=N_{2}=4$ for the Fock space dimensions of each cavity field. When we increase the Fock space dimension up to 8 and remake our calculations, we find that the results remain unchanged. We consider two cases distinguished by the action of weak $F/\kappa=0.1$ and strong pump $F/\kappa=1$. We take $\kappa/2\pi=0.4$ MHz. The parameters we use in the simulations are within the ranges accessible in present circuit QED systems \cite{PhysRevLett.105.140502,PhysRevA.75.032329}.

Numerically computed steady state density operator is used for calculation of second order coherence, spin squeezing, concurrence and mode entanglement parameters. Zero-time delay second order quantum coherence function is defined by~\cite{scully1997quantum}
\begin{eqnarray}\label{eq:coherence}
g^{(2)}(0)&=&\frac{\tr(\hat{b}^{\dagger}\hat{b}^{\dagger}\hat{b}\hat{b}\hat{\rho})}{[\tr(\hat{b}^{\dagger}\hat{b}\hat{\rho})]^2},\\
&=&1+\frac{\langle(\Delta \hat n)^2\rangle-\langle \hat n\rangle}{\langle \hat n\rangle^2},
\end{eqnarray}
Here $\hat n=\hat b^\dag \hat b$, $\Delta\hat n=\hat n-\langle \hat n\rangle$; and we use $\rho$ instead of $\rho^{(j)}$ for notational simplicity. We evaluate $g^{(2)}(0)$ for $\hat{b}=\hat{b}_{i}$. Both cavity photons have identical coherence functions due to exchange symmetry of our models under $b_1\leftrightarrow b_2$. $g^{(2)}(0)<1$ implies sub-Poissonian statistics of photons, while $g^{(2)}(0)>1$ implies super-Poissonian statistics. Coherent photons are recognized by $g^{(2)}(0)=1$. Sub-Poissonian statistics is a violation of
the Cauchy-Schwartz inequality obeyed by classical light; and hence it is a profound manifestation of quantum light.

Spin squeezing is witnessed by the inequality~\cite{PhysRevLett.99.250405}
\begin{equation}\label{eq:sq}
\langle \hat{J}_{k}^2\rangle+\langle \hat{J}_{l}^2\rangle-\frac{N}{2}\leq(N-1)(\Delta \hat{J}_{m})^2,
\end{equation}
where $k,l,m$ take all the possible permutations of $x,y,z$. Inequality given in Eq.~(\ref{eq:sq}) is one of the four optimal spin squeezing inequalities introduced in Ref.~\cite{PhysRevLett.99.250405} with $N$ being the total number of particles in the system. Violation of it implies spin squeezing and particle entanglement and its relation to the positivity of the concurrence can be found in Ref.~\cite{vidal2006concurrence}. We rewrite the optimal spin squeezing inequality given in Eq.~(\ref{eq:sq}) with $k=x,l=z$ and $m=y$ as
\begin{equation}\label{eq:sq1}
\zeta\equiv\langle \hat{J}_{x}^2\rangle+\langle \hat{J}_{z}^2\rangle-\frac{N}{2}-(N-1)(\Delta \hat{J}_{y})^2\leq0.
\end{equation}
The positive values of $\zeta$ indicate spin squeezing and multi-particle entanglement in the first quantization description.
As we have an open system, $N=\langle \hat N\rangle=\langle \hat{n}_{1}\rangle+\langle \hat{n}_{2}\rangle$, with $\hat N=\hat n_1+\hat n_2$ and $\hat n_i=\hat b_i^\dag \hat b_i$, is not conserved and  varies in time.

\begin{figure*}
     \begin{center}
        \subfigure[]{%
            \label{fig:fig2a}
            \includegraphics[width=4cm]{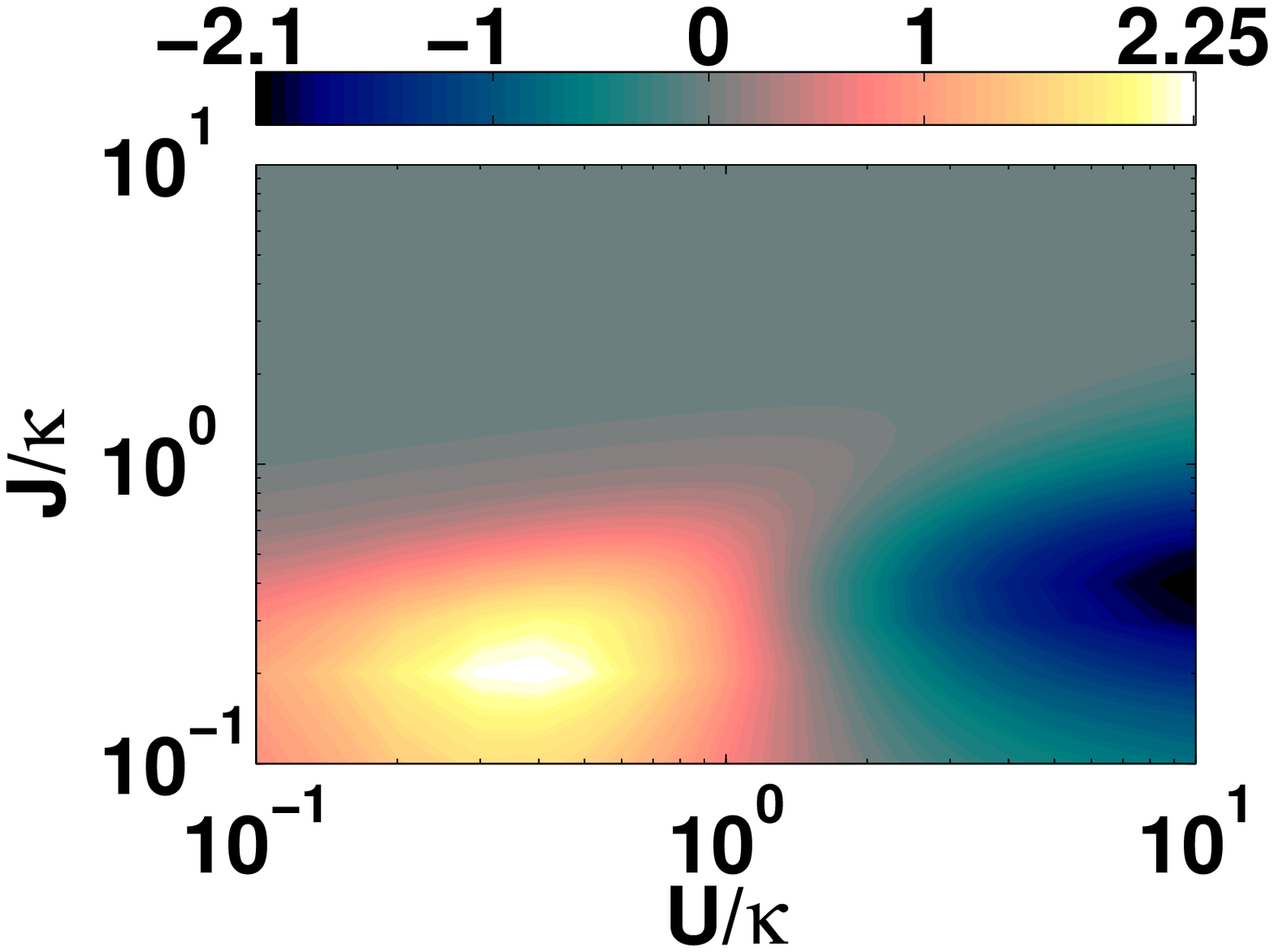}
        }%
        \subfigure[]{%
           \label{fig:fig2b}
           \includegraphics[width=4cm]{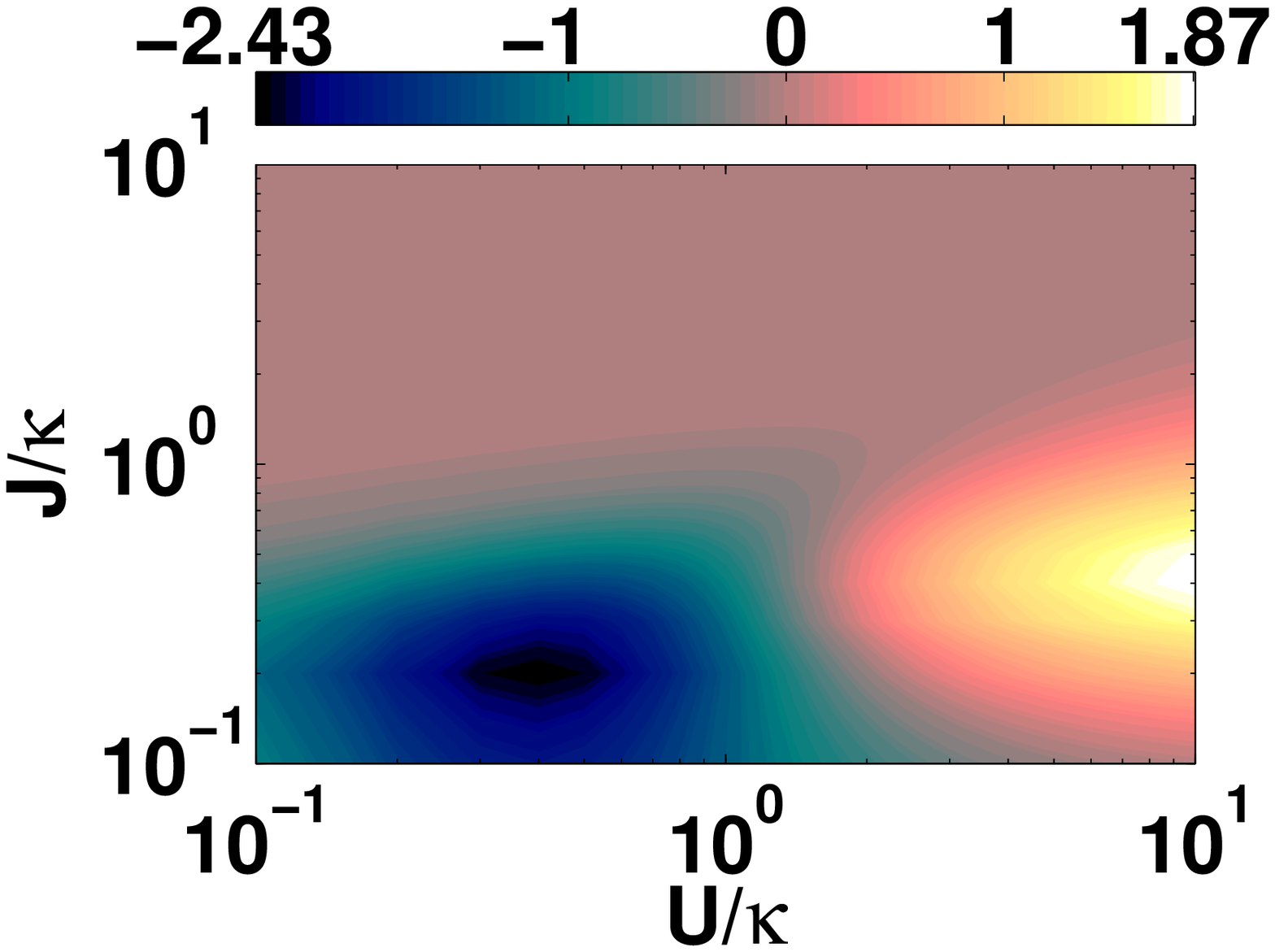}
        }%
           \subfigure[]{%
            \label{fig:fig2c}
            \includegraphics[width=4cm]{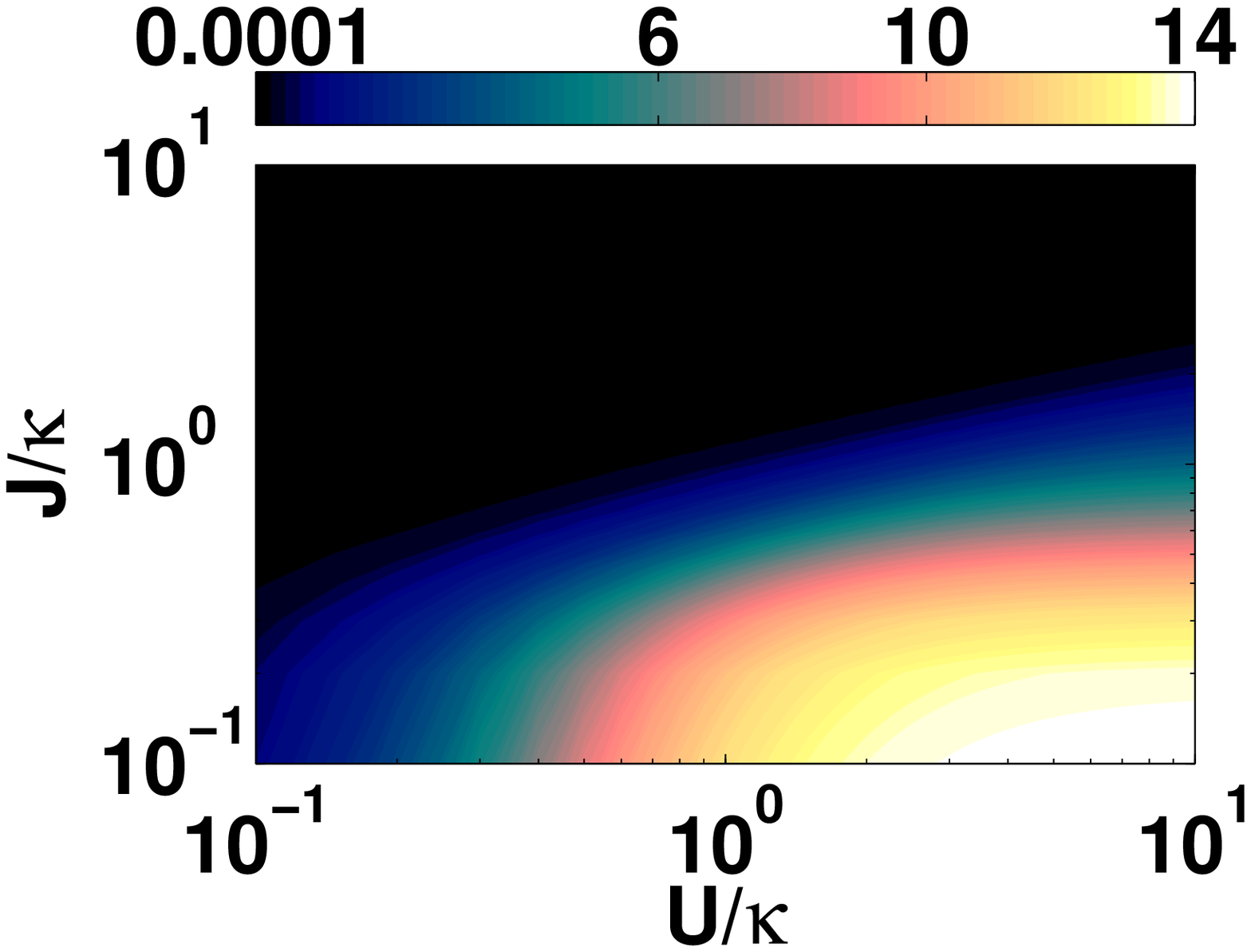}
         }%
           \subfigure[]{%
            \label{fig:fig2d}
            \includegraphics[width=4cm]{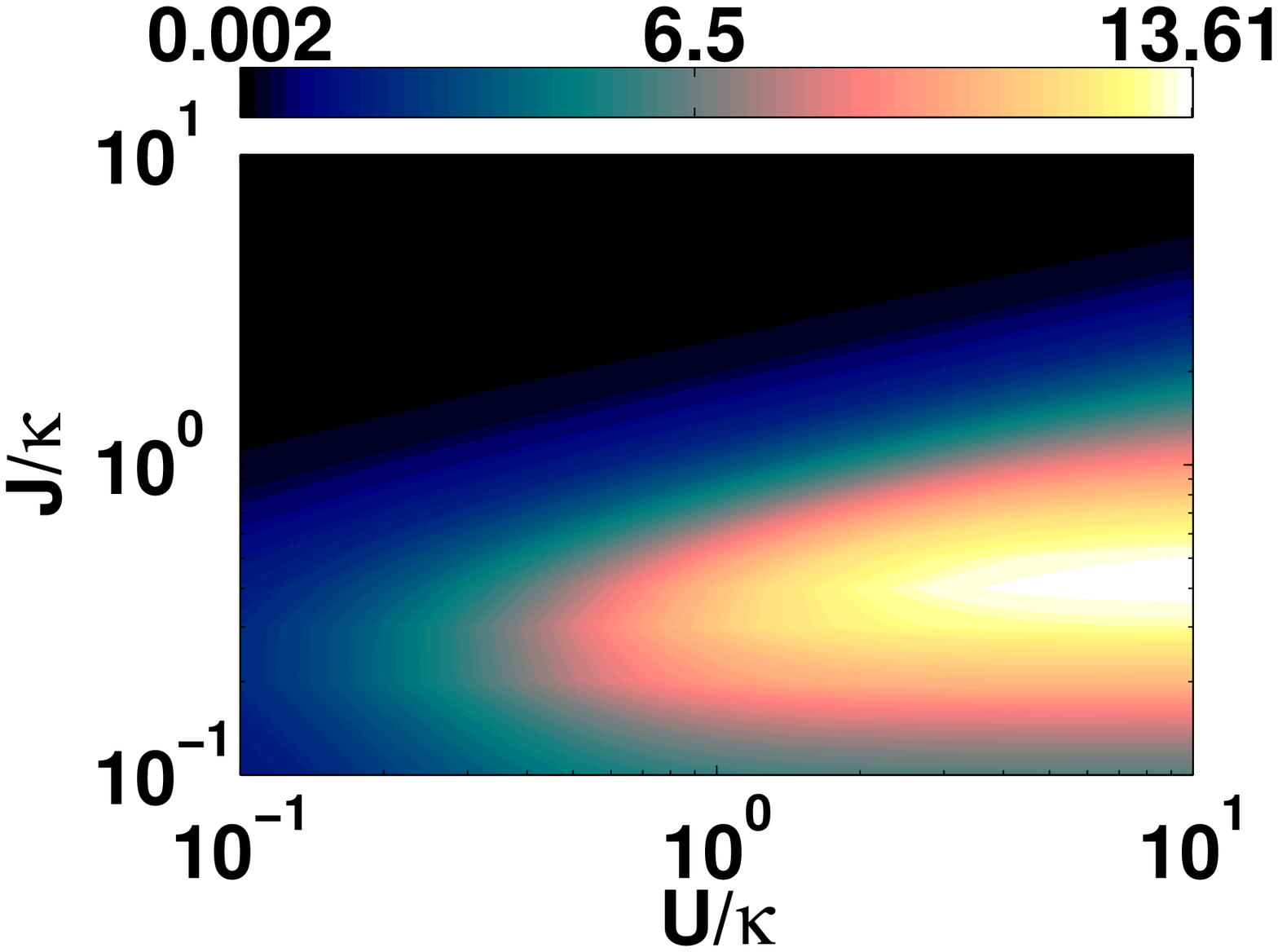}    
        }\\ 
        \subfigure[]{%
            \label{fig:fig2e}
            \includegraphics[width=4cm]{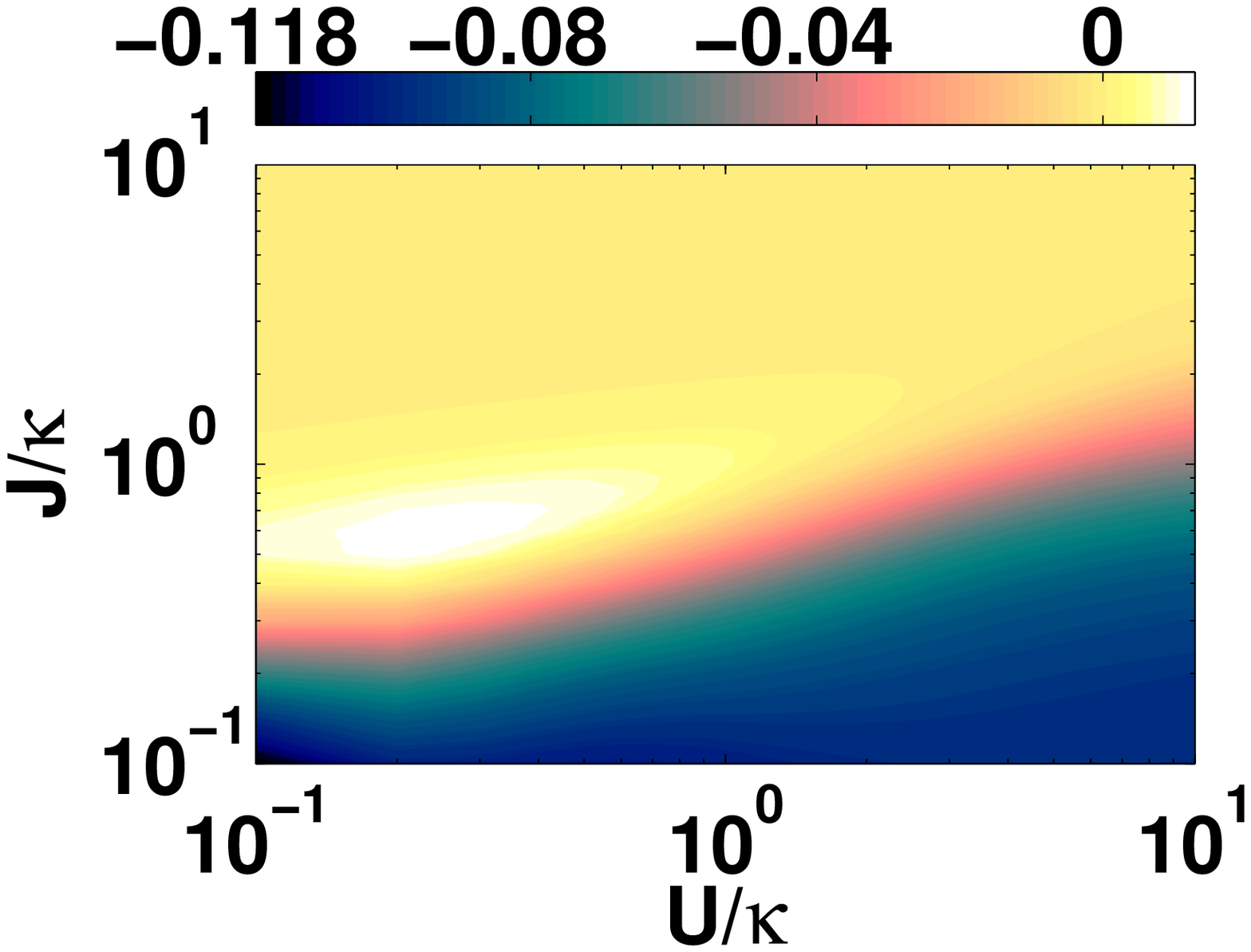}
        }%
        \subfigure[]{%
            \label{fig:fig2f}
            \includegraphics[width=4cm]{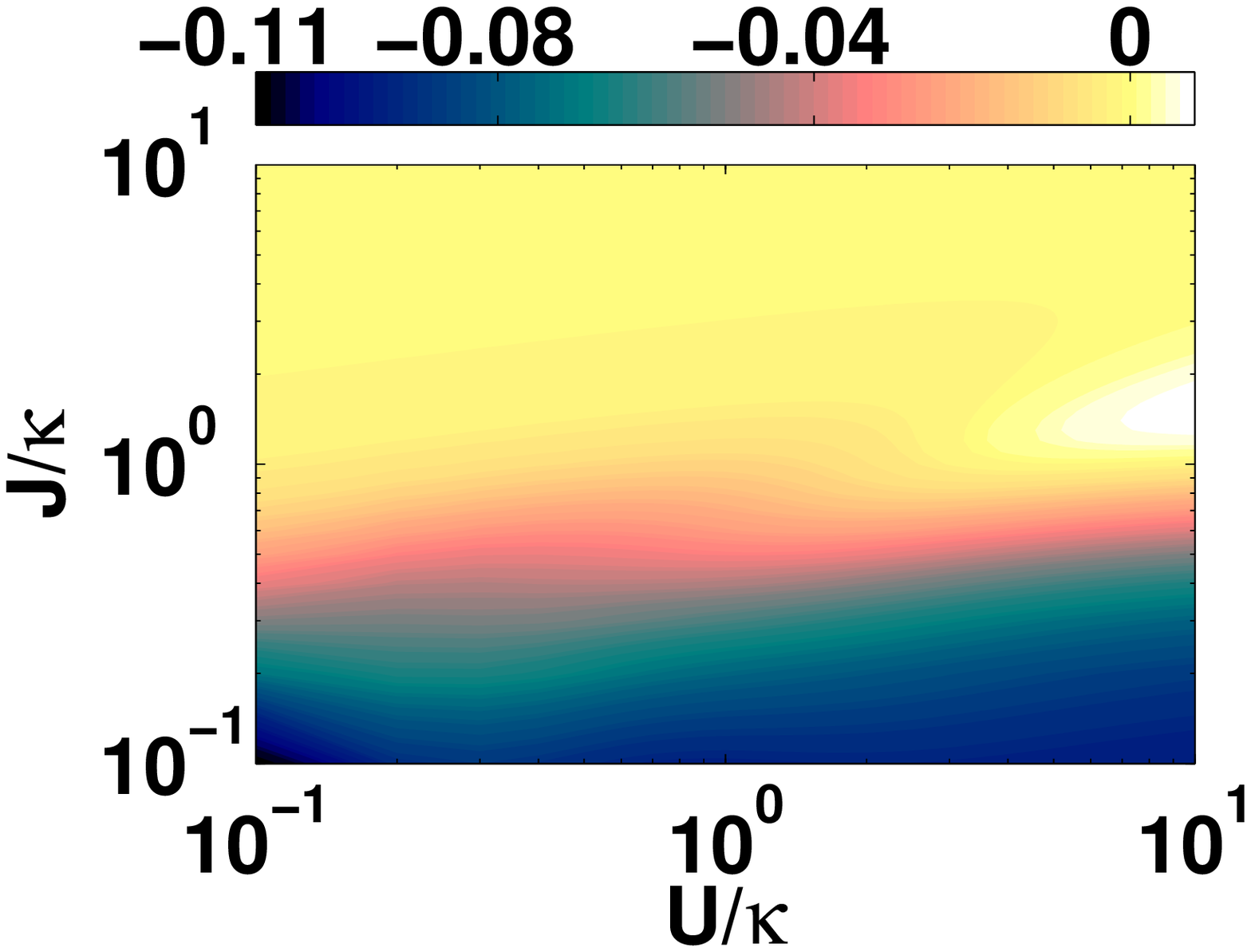}
        }%
        \subfigure[]{%
            \label{fig:fig2g}
            \includegraphics[width=4cm]{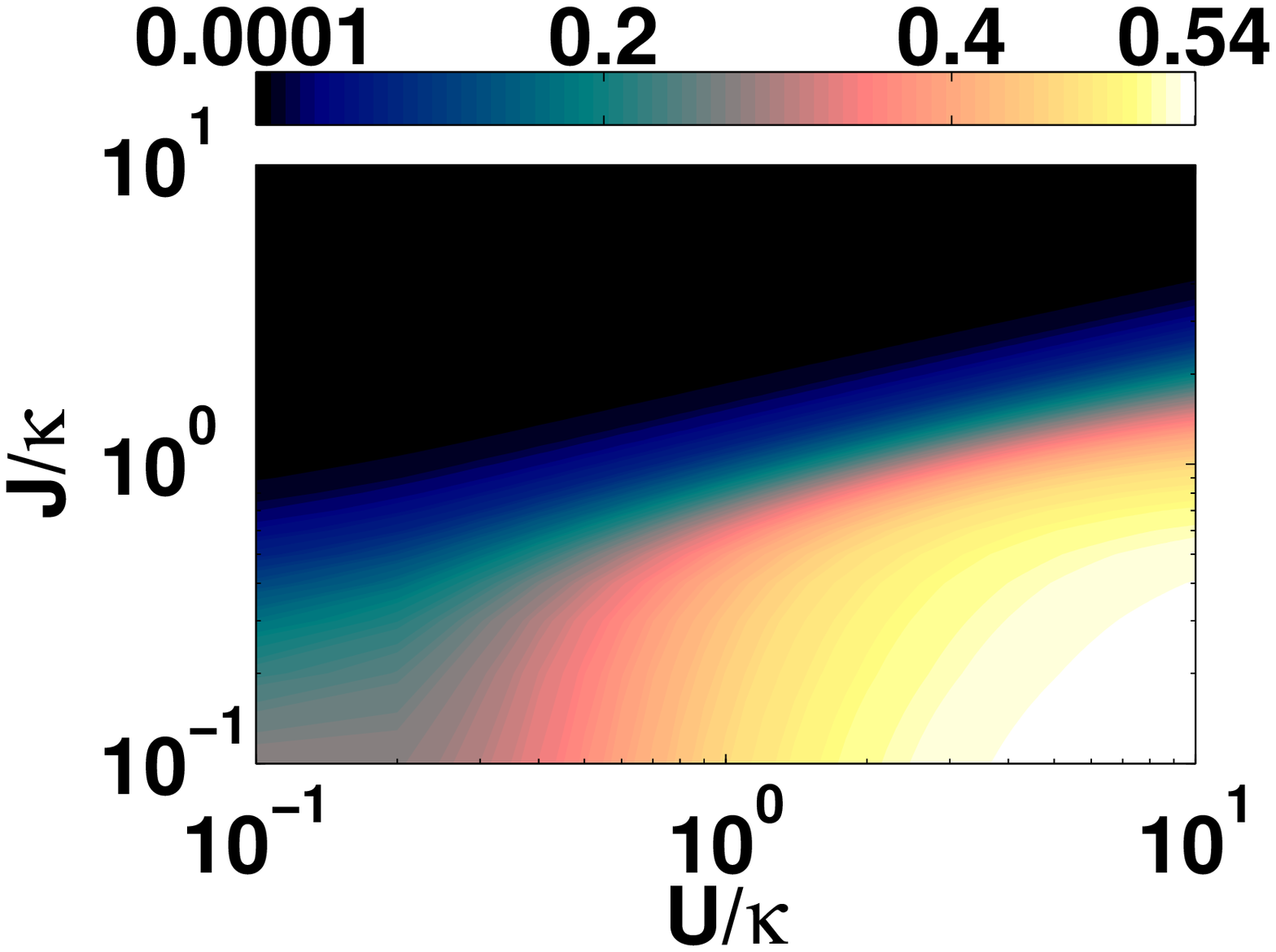}
	}%
        \subfigure[]{%
            \label{fig:fig2h}
            \includegraphics[width=4cm]{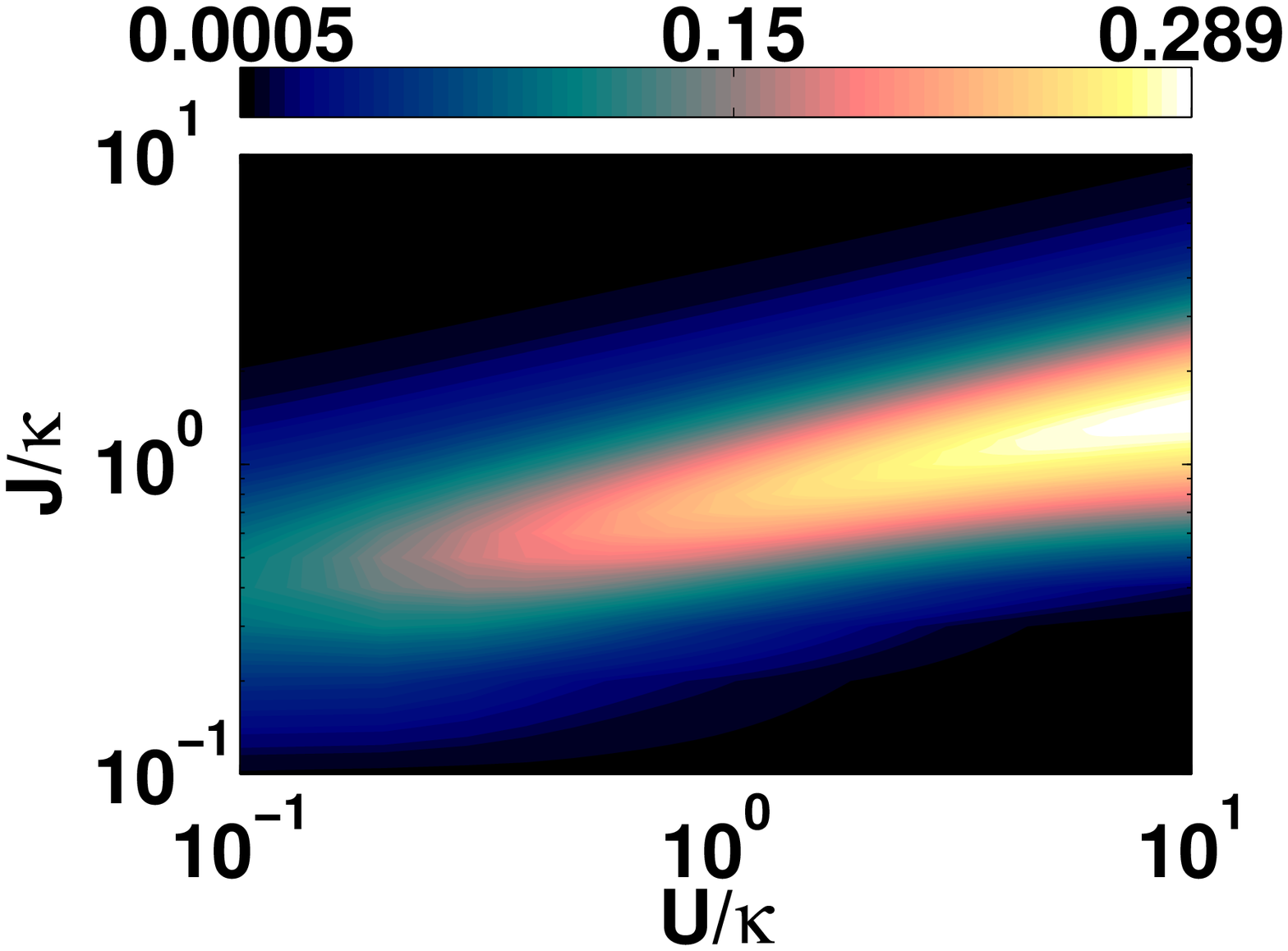}
        }%
    \end{center}
    \caption{%
        (Color Online) Dependence of (a) $\lambda_{1}$, (b) $\lambda_{2}$ (c) $S(\rho_{1})$ for $F/\kappa=0.1$ (d) $E_N(\rho)$ and (e) $\lambda_{1}$, (f) $\lambda_{2}$ (g) $S(\rho_{1})$ (h) $E_N(\rho)$ for $F/\kappa=1$ with respect to dimensionless $J/\kappa$ and $U/\kappa$ in two-site KH system with single-photon exchange. $\lambda_{1}$ and $\lambda_{2}$ are multiplied by $10^5$, $S(\rho_{1})$ is multiplied by $10^4$, and $E_N(\rho)$ is multiplied by $10^{3}$ in the case of $F/\kappa=0.1$ for visibility.
     }%
   \label{fig:fig2}
\end{figure*}

In addition, we analyze pairwise particle entanglement through so called I-concurrence~\cite{PhysRevA.64.042315} given as
\begin{equation}\label{eq:concur}
C_{I}=\sqrt{2(1-\tr{\rho_{i}^2})},
\end{equation}
where $\rho_{i}$ with $i=1,2$ is the reduced density operator of the subsystem under consideration. $C_I=0$ for product states and nonzero for entangled states. The maximum value of $C_{I}$ cannot exceed $C_{I}^{max}=\sqrt{2(1-1/N)}$ where $N=4$ is the Hilbert space dimension of the subsystem. For our case it is found to be $C_{I}^{max}\cong1.225$.

We investigate the entanglement in the second quantized description in terms of the genuine two-mode entanglement parameters $\lambda_{1}$ and $\lambda_{2}$ given by~\cite{PhysRevLett.96.050503,PhysRevA.74.032333}
\begin{eqnarray}\label{eq:mode}
\lambda_{1}&=&|\langle\hat{b}_{1}^{\dagger}\hat{b}_{2}\rangle|^2-\langle\hat{n}_{1}\hat{n}_{2}\rangle,\\
\lambda_{2}&=&|\langle\hat{b}_{1}\hat{b}_{2}\rangle|^2-\langle\hat{n}_{1}\rangle\langle\hat{n}_{2}\rangle,
\end{eqnarray}
where $\hat{n}_{i}=\hat{b}_{i}^{\dagger}\hat{b}_{i}$ with $i=1,2$. Positivity of the parameters $\lambda_{1}$ and  $\lambda_{2}$ indicates mode entanglement. Positivity of the $\lambda_{1}$ and $\lambda_{2}$ is an only sufficient condition. On the other hand, their usability for the investigation of modal entanglement has recently been established~\cite{PhysRevA.87.022325}.

The von Neumann entropy for the reduced density operator $\rho_{i}$ with $i=1,2$ is given by~\cite{nielsen2010quantum}
\begin{equation}\label{eq:entropy}
S(\rho_{i})=-\tr{\rho_{i}\ln{\rho_{i}}}.
\end{equation}
The maximum value of the entropy is given by $S(\rho_{i})^{max}=\ln N=\ln4\cong1.386$. Its nonzero values detect the entanglement and is a measure for bipartite entanglement. 

I-Concurrence and von Neumann entropy are reliable to characterize pairwise and bipartite entanglement in pure states. Steady states of our models may not be pure under strongly nonlinearity and strong drive. We calculate the impurity parameter $I=1-Tr(\rho^2)$ and report the plots of logarithmic negativity $E_N(\rho)$, which is applicable to mixed state bipartite entanglement. 
It is defined as~\cite{vidal02}
\begin{eqnarray}
E_N(\rho)=\log_2\vert\vert\rho^{T_1}\vert\vert_1,
\end{eqnarray}
where $\rho^{T_1}$ is the partial transpose with respect to the first subsystem $i=1$ and $\vert\vert\rho^{T_1}\vert\vert_1$ is the trace norm of $\rho^{T_1}$. Nonlinearity in our system and large Hilbert space dimensions make the calculation of other mixed state entanglement measures, which are convenient for Gaussian states or states of finite dimensions, too difficult while $E_N(\rho)$ is a computable and non-convex entanglement monotone~\cite{plenio05}.
\begin{figure*}
     \begin{center}
        \subfigure[]{%
            \label{fig:fig3a}
            \includegraphics[width=4cm]{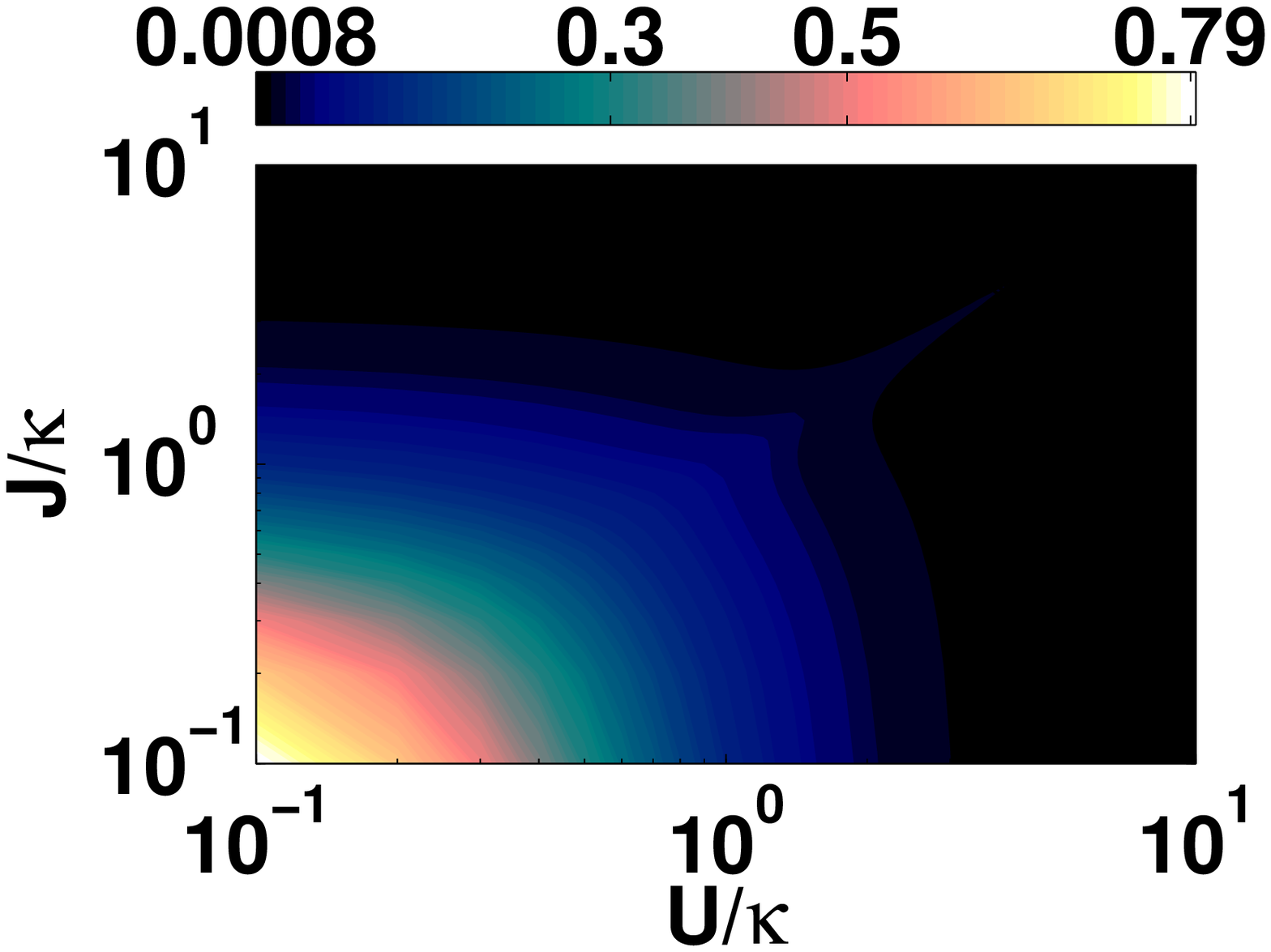}
        }%
        \subfigure[]{%
           \label{fig:fig3b}
           \includegraphics[width=4cm]{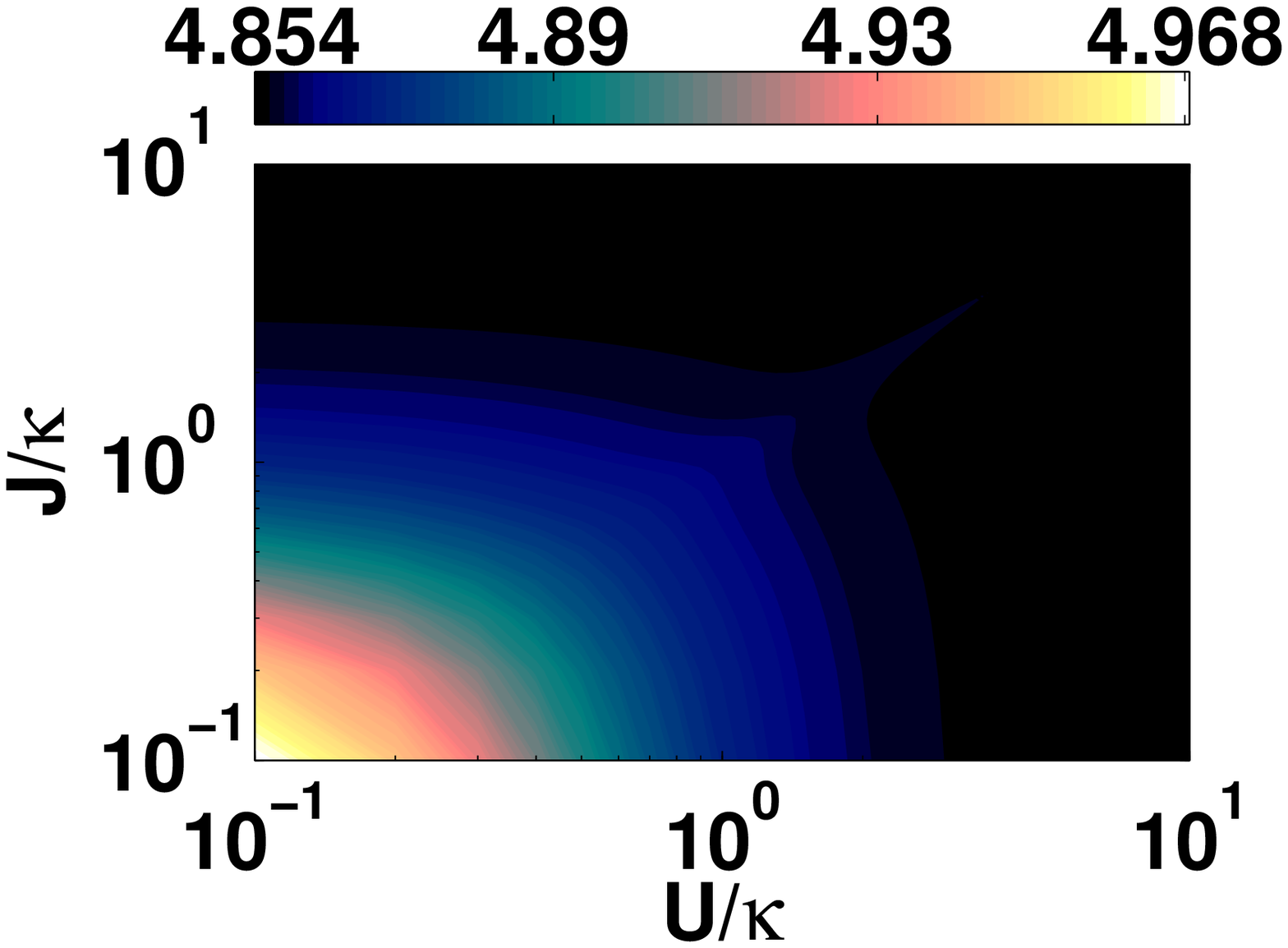}
        }%
           \subfigure[]{%
            \label{fig:fig3c}
            \includegraphics[width=4cm]{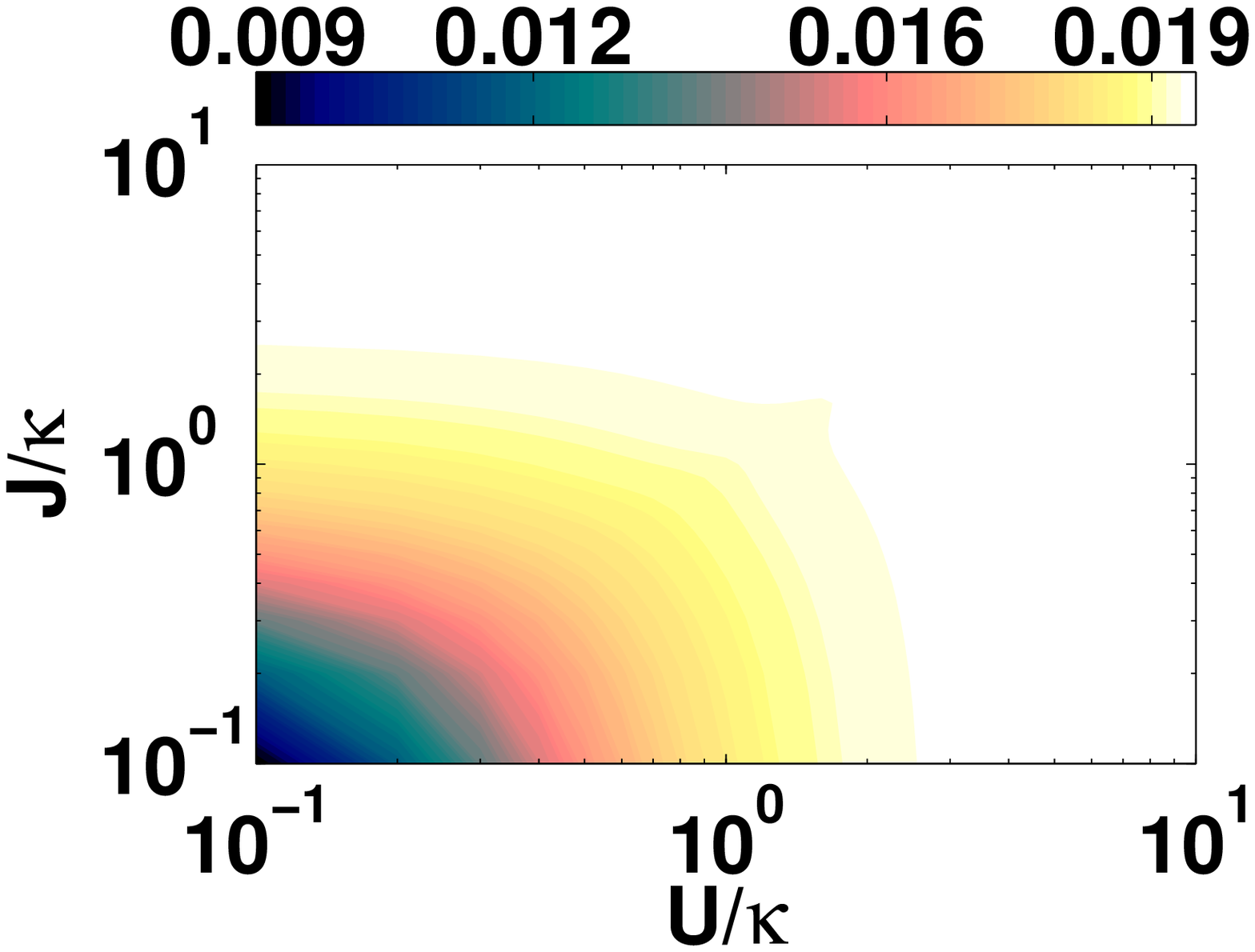}
        }\\ 
        \subfigure[]{%
            \label{fig:fig3d}
            \includegraphics[width=4cm]{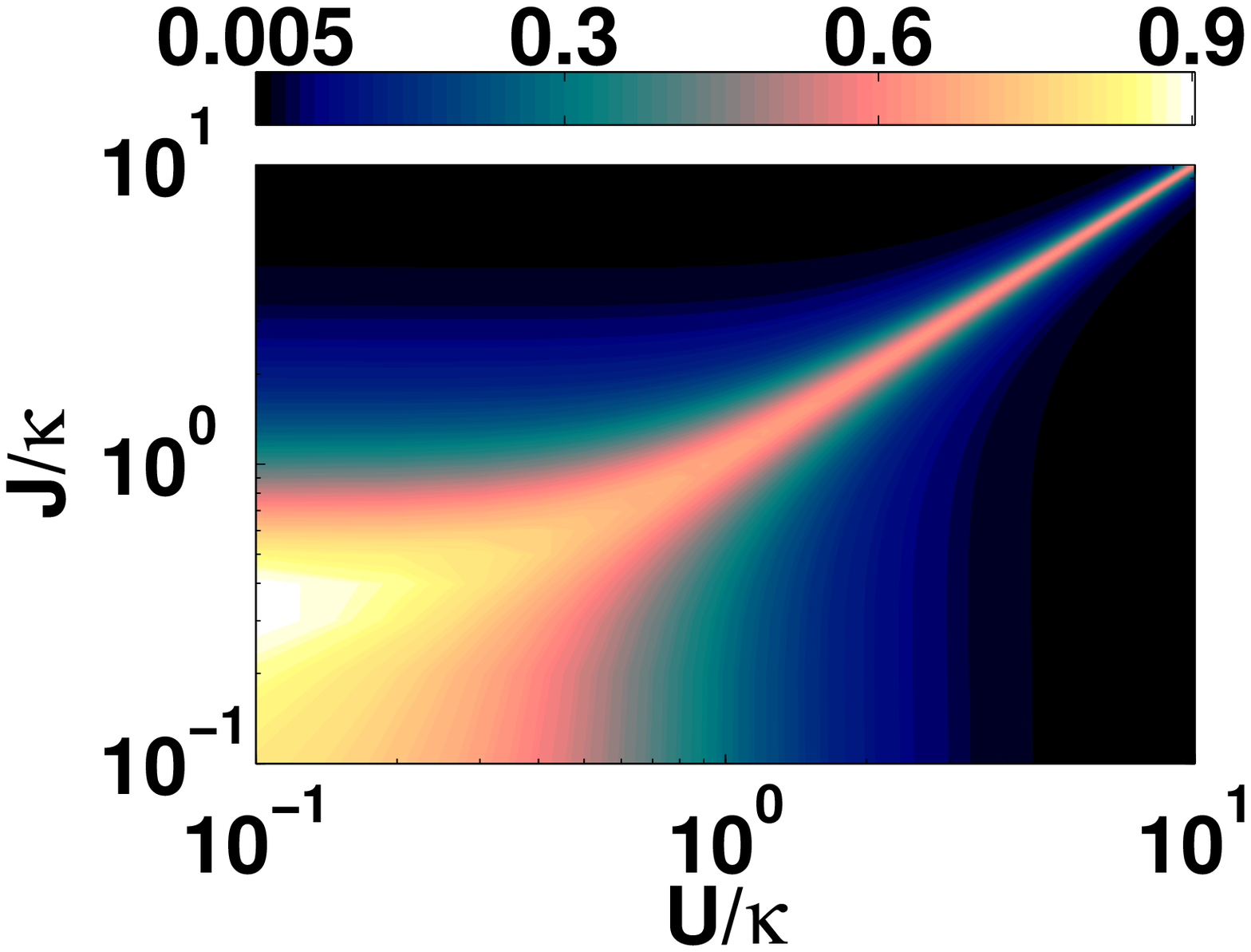}
        }%
        \subfigure[]{%
            \label{fig:fig3e}
            \includegraphics[width=4cm]{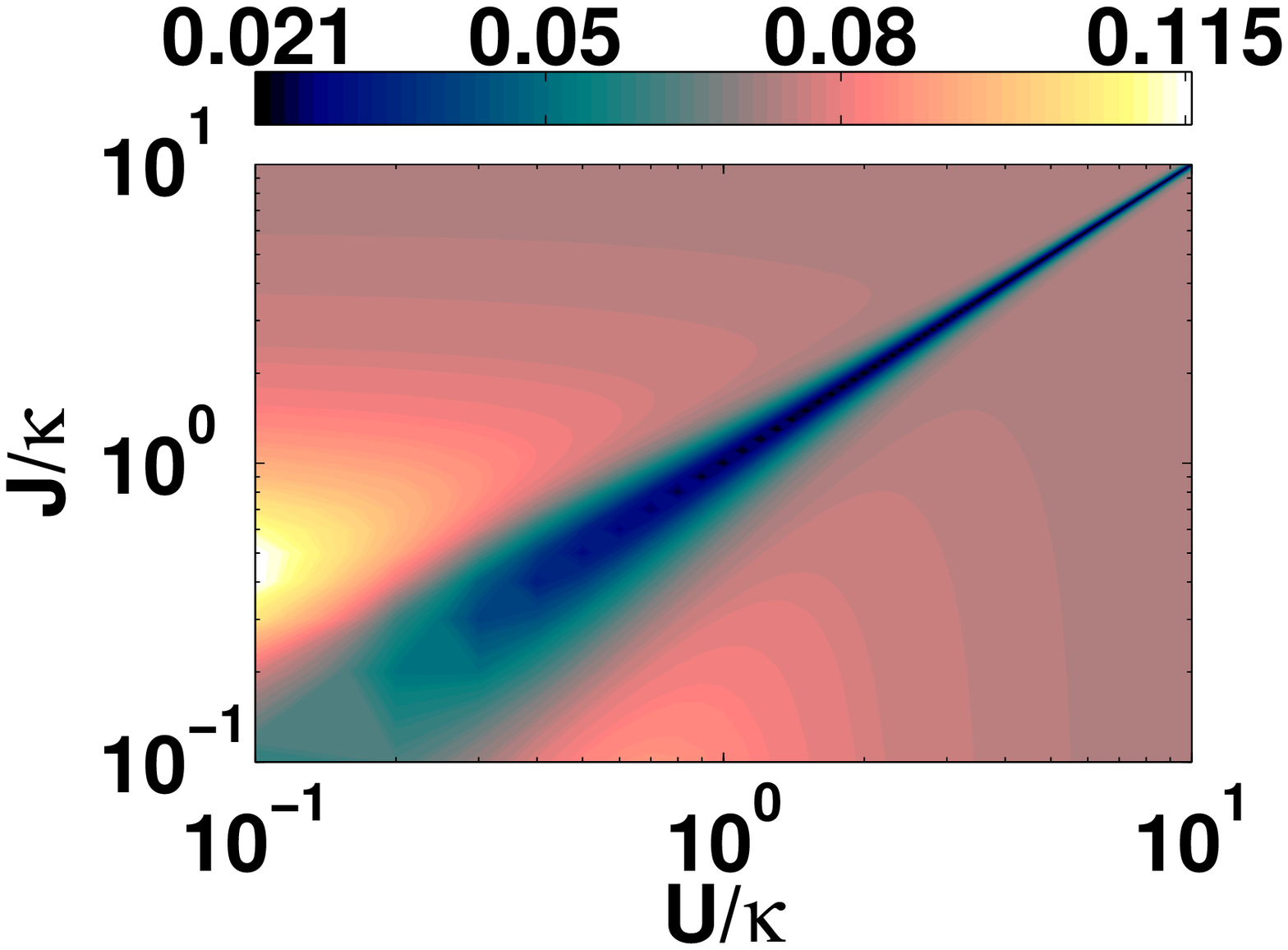}
        }%
        \subfigure[]{%
            \label{fig:fig3f}
            \includegraphics[width=4cm]{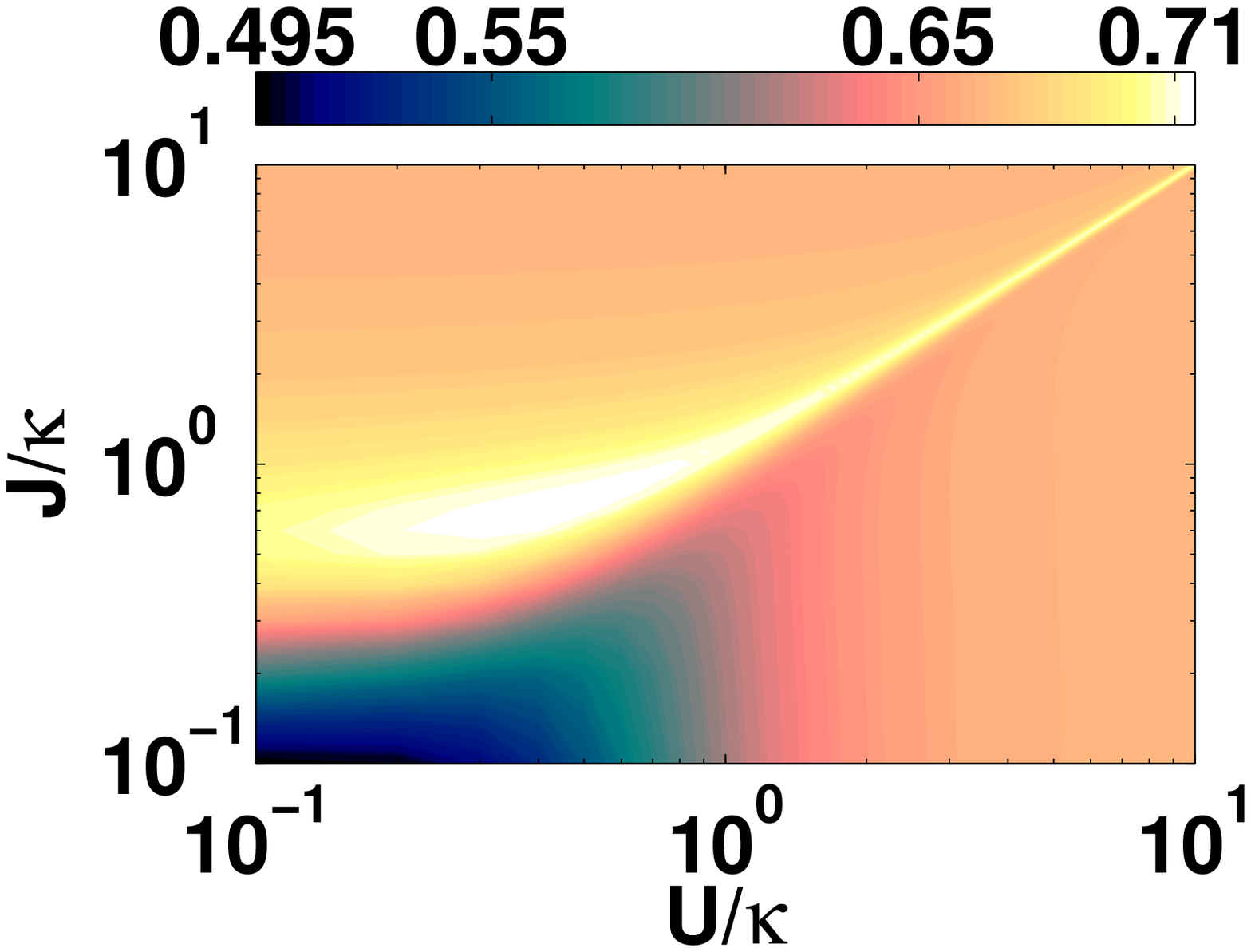}
        }%
    \end{center}
    \caption{%
        (Color Online) Dependence of (a) $g^{(2)}(0)$, (b) $\zeta$ (c) $C_{I}$ for $F/\kappa=0.1$ and (d) $g^{(2)}(0)$, (e) $\zeta$ (f) $C_{I}$ for $F/\kappa=1$ with respect to dimensionless $J/\kappa$ and $U/\kappa$ in two-site KH system with two-photon exchange. $\zeta$ is multiplied by $10^3$ in the case of $F/\kappa=0.1$ for visibility.
     }%
   \label{fig:fig3}
\end{figure*}

An intuitive  link between coherence, particle and mode entanglements can be formed by considering that coherence function is a relative measure of number fluctuations which carries information on deviation from Poisson statistics.
One can establish an analogy between the number operator $\hat{n}$ and the z-component of the angular momentum operator $\hat{J}_{z}$ using $\hat n_1=\hat N / 2 + \hat J_z$. The coherence function is influenced by squeezing the spin noise around the $z$ axis. Under stringent conditions spin squeezing implies particle entanglement. Particle like behavior, enforced by
the sub-Poissonian statistics can be associated with spin squeezing type pairwise nonlinear interactions.
In that case coherence and spin squeezing interplay can be further extended to multiparticle entanglement.
Similarly $\lambda_{1}$ and $\lambda_{2}$ can be expressed in terms of spin
fluctuations and thus would be influenced by the coherence and spin squeezing in the system.

Above heuristic arguments motivates the existence of an interplay between coherence, particle and mode entanglement. Let us now analyze specific
cases numerically. We will first consider the single photon exchange case, then proceed to two photon exchange case in the following subsections.
\subsection{Single-Photon Exchange}\label{sec:single}
In Fig.~\ref{fig:fig1a}, Fig.~\ref{fig:fig1b} and Fig.~\ref{fig:fig1c} we plot second order coherence, spin squeezing and concurrence parameters as functions of nonlinearity and photon exchange coefficients in the case of weak drive with $F/2\pi=0.04$ MHz, respectively. Fig.~\ref{fig:fig1a} reproduces the result in Ref.~\cite{ferretti2010photon}. Second order coherence varies over the range of $0.00183\leq g^{(2)}(0)\leq1.011$. Coherent delocalization of cavity photons happens for $J/\kappa>1$ and $J>U$, which corresponds to the Poissonian statistics indicated by $g^{(2)}(0)\sim 1$. Strong sub-Poissonian localization regime with $g^{(2)}(0)\sim 0$ lies in the region
of $U/\kappa>1$ and $J/\kappa<1$.

Fig.~\ref{fig:fig1b} shows that spin squeezing always present in the system for the ranges of $0.1<U/\kappa,J/\kappa<10$. Pairwise nonlinearity measured by $U$ in the driven system is strong enough for the survival of multiparticle entanglement in the steady state. It is easier to violate spin squeezing inequality in the region of $J/\kappa<1$, where sub-Poissonian and hence particle like behavior is more significant.
This is intuitively expected as the delocalizing photon hopping makes particle entanglement more difficult, while nonlinear interaction favors particle correlations.

Fig.~\ref{fig:fig1c} shows that the pairwise entanglement characterized with concurrence is in agreement with the
spin squeezing. It reaches its maximal value in the strong Sub-Poissonian regimes, where particle entanglement is maximal.

Fig.~\ref{fig:fig1d}, Fig.~\ref{fig:fig1e} and Fig.~\ref{fig:fig1f} depict quantum coherence, spin squeezing and concurrence for the case of strong drive of $F/2\pi=0.4$ MHz, respectively. Coherent effects of the drive is harmful for the particle like character and the range of $g^{(2)}(0)$ is shifted away from $0$ and becomes $0.00486\leq g^{(2)}(0)\leq1.0321$. Stronger nonlinearity is needed for sub-Poissonian statistics and thus the region is narrowed in the $U/\kappa$ axis relative to that in Fig.~\ref{fig:fig1a}.
 Multiparticle and pairwise entanglement becomes harder to establish relative to weak drive case. The maximum values of $\zeta$ and $C_{I}$ reduce about an order of magnitude when the coherent drive is increased an order of magnitude. Spin squeezing and concurrence are almost absent in the super-Poissonian region.
\begin{figure*}
     \begin{center}
        \subfigure[]{%
            \label{fig:fig4a}
            \includegraphics[width=4cm]{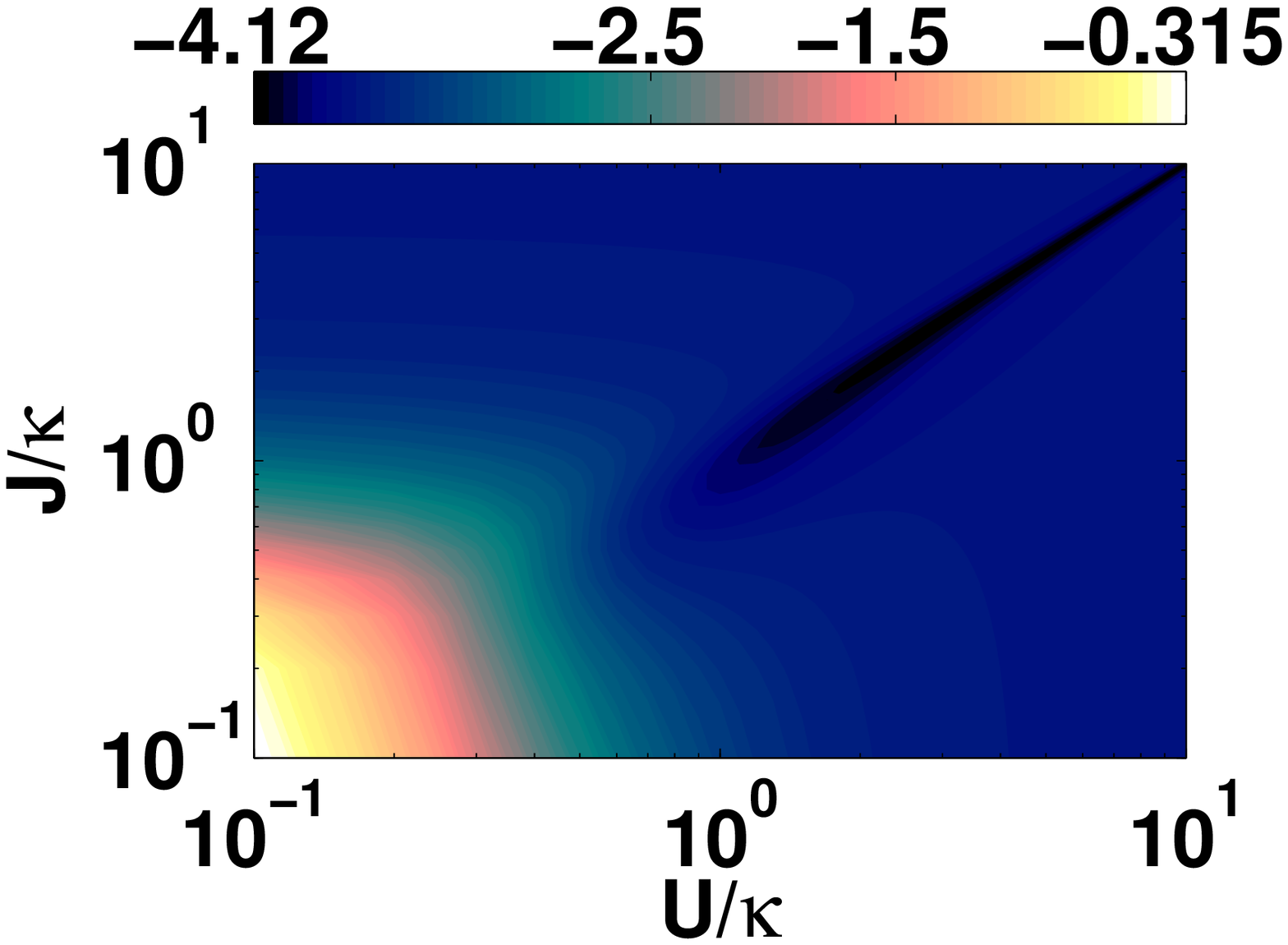}
        }%
        \subfigure[]{%
           \label{fig:fig4b}
           \includegraphics[width=4cm]{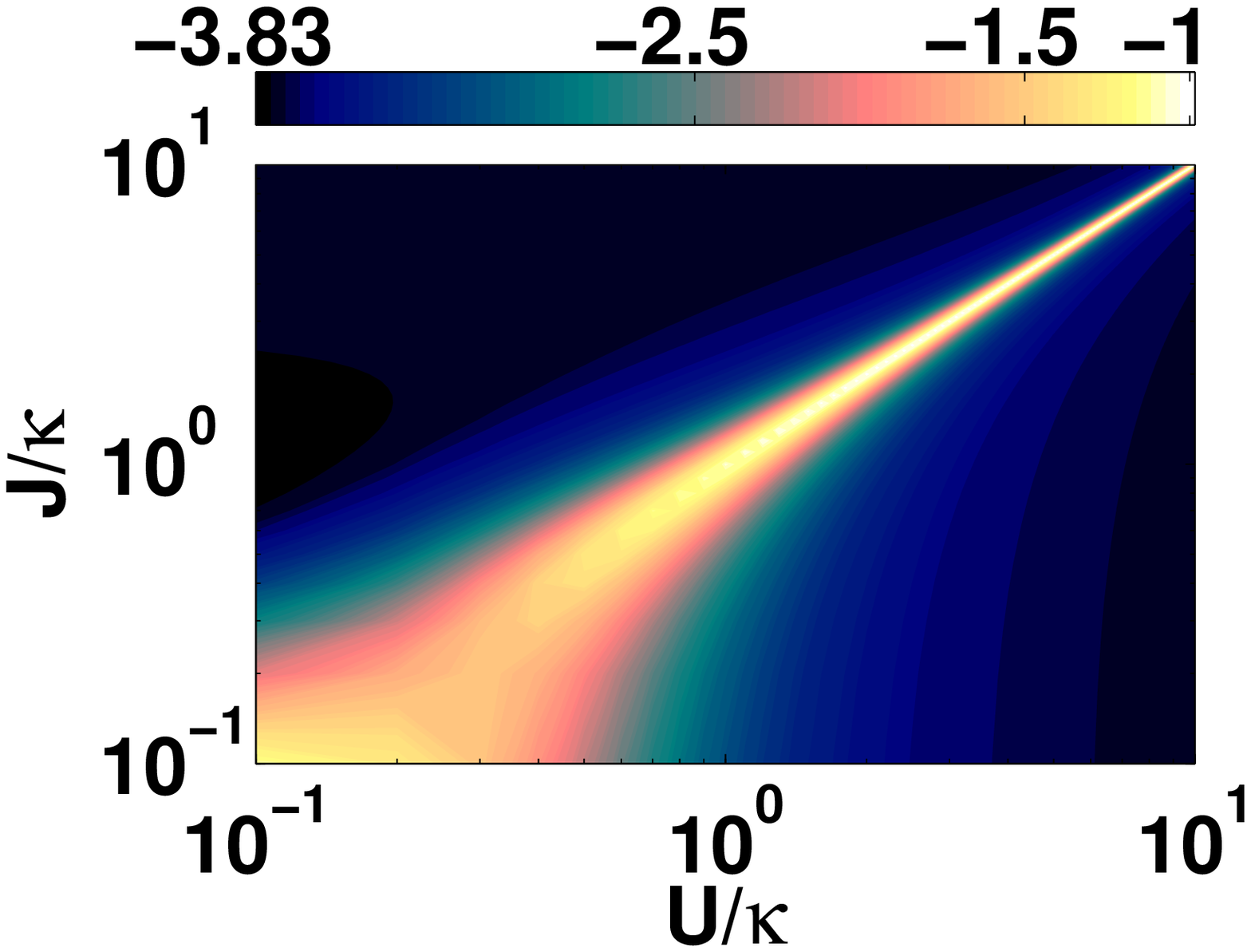}
        }%
           \subfigure[]{%
            \label{fig:fig4c}
            \includegraphics[width=4cm]{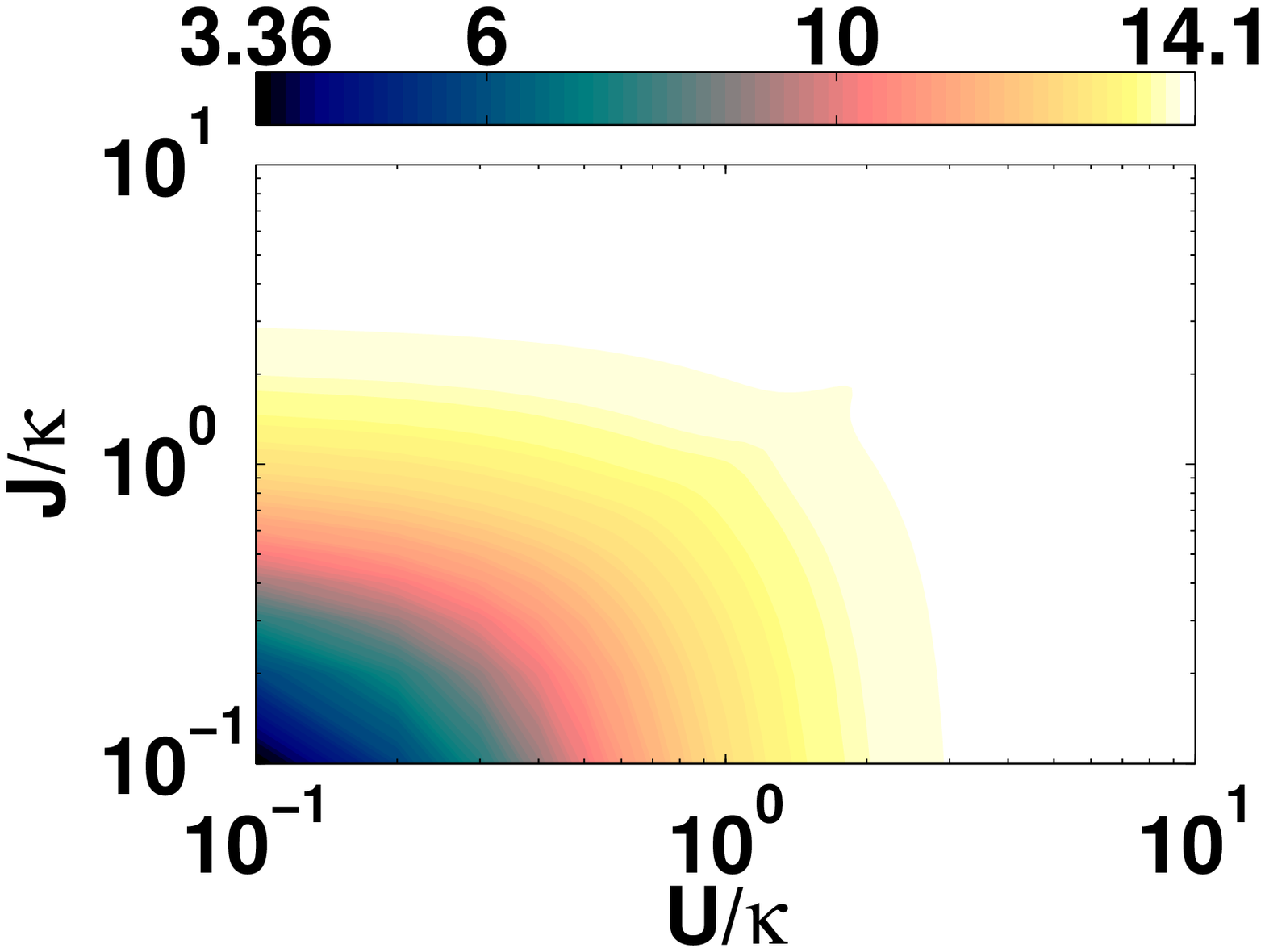}
        }%
           \subfigure[]{%
            \label{fig:fig4d}
            \includegraphics[width=4cm]{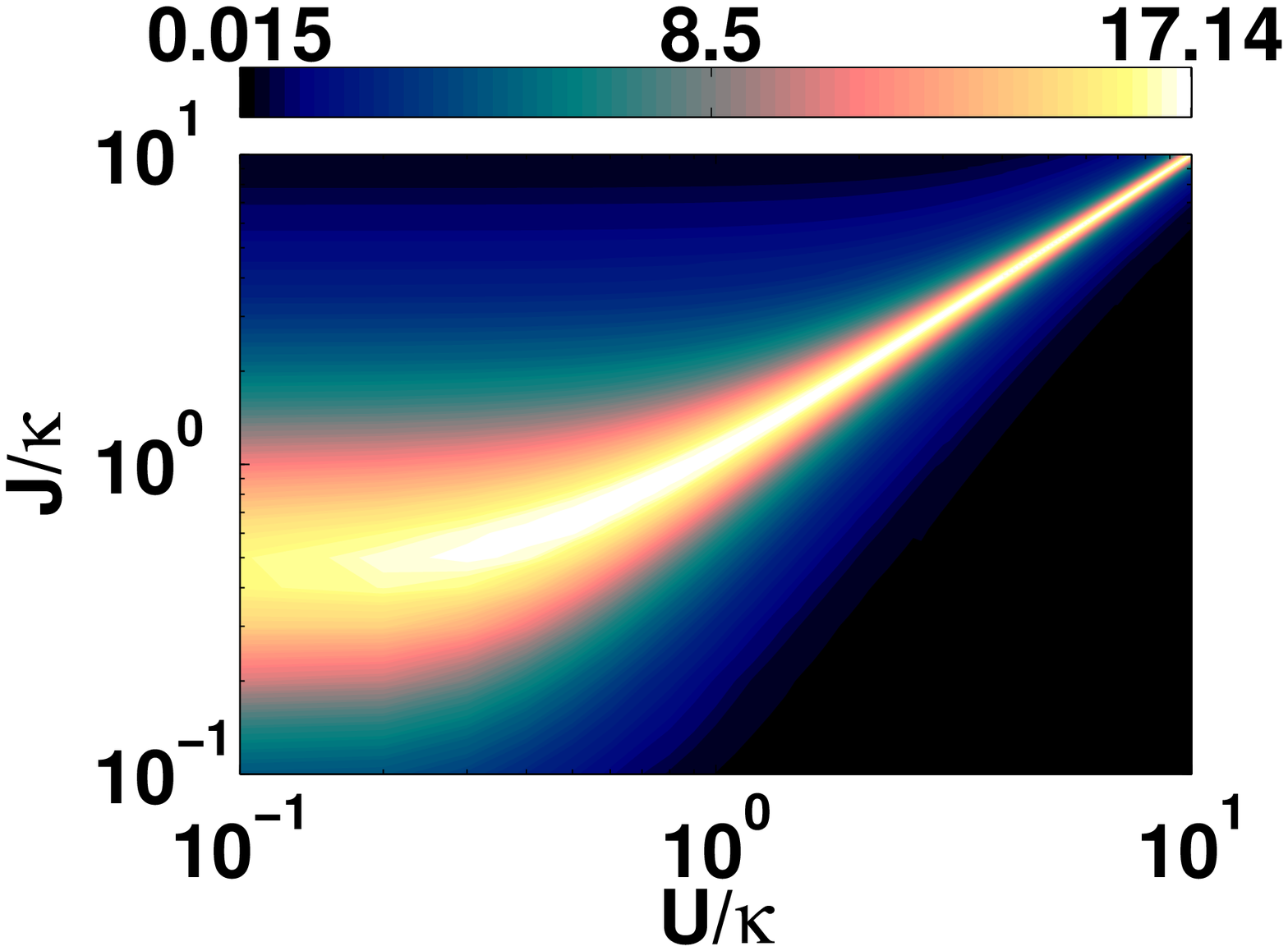}
        }\\ 
        \subfigure[]{%
            \label{fig:fig4e}
            \includegraphics[width=4cm]{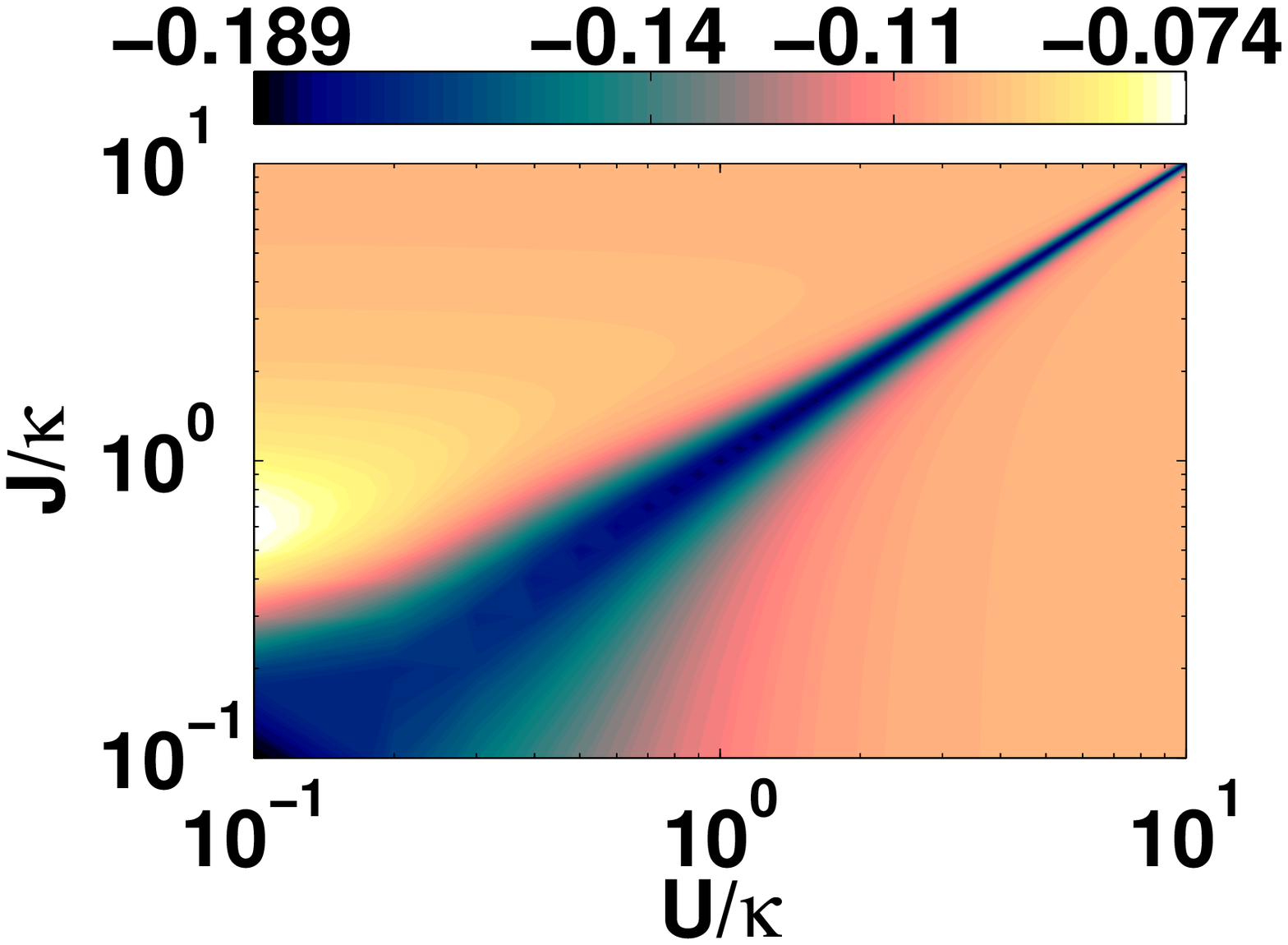}
        }%
        \subfigure[]{%
            \label{fig:fig4f}
            \includegraphics[width=4cm]{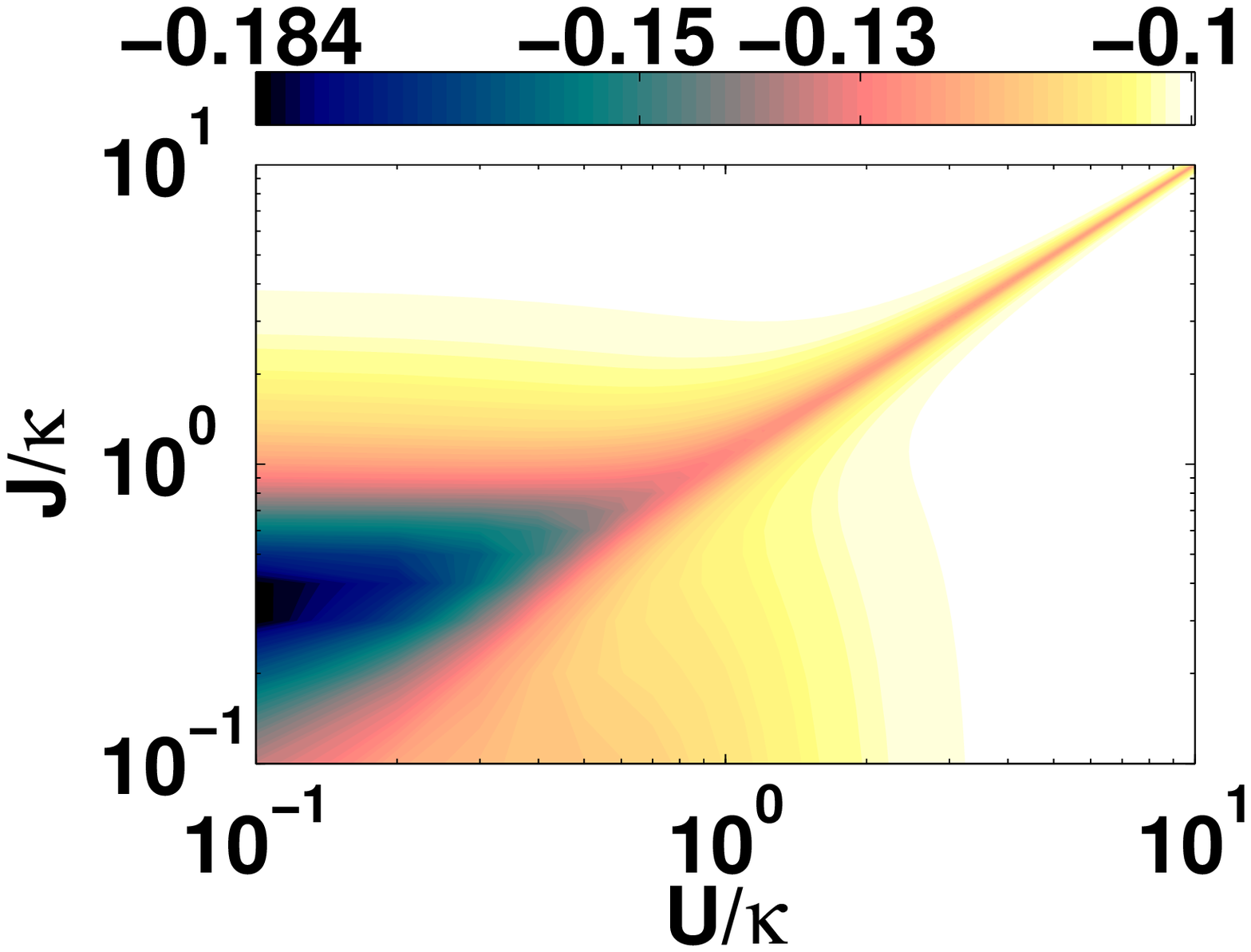}
        }%
        \subfigure[]{%
            \label{fig:fig4g}
            \includegraphics[width=4cm]{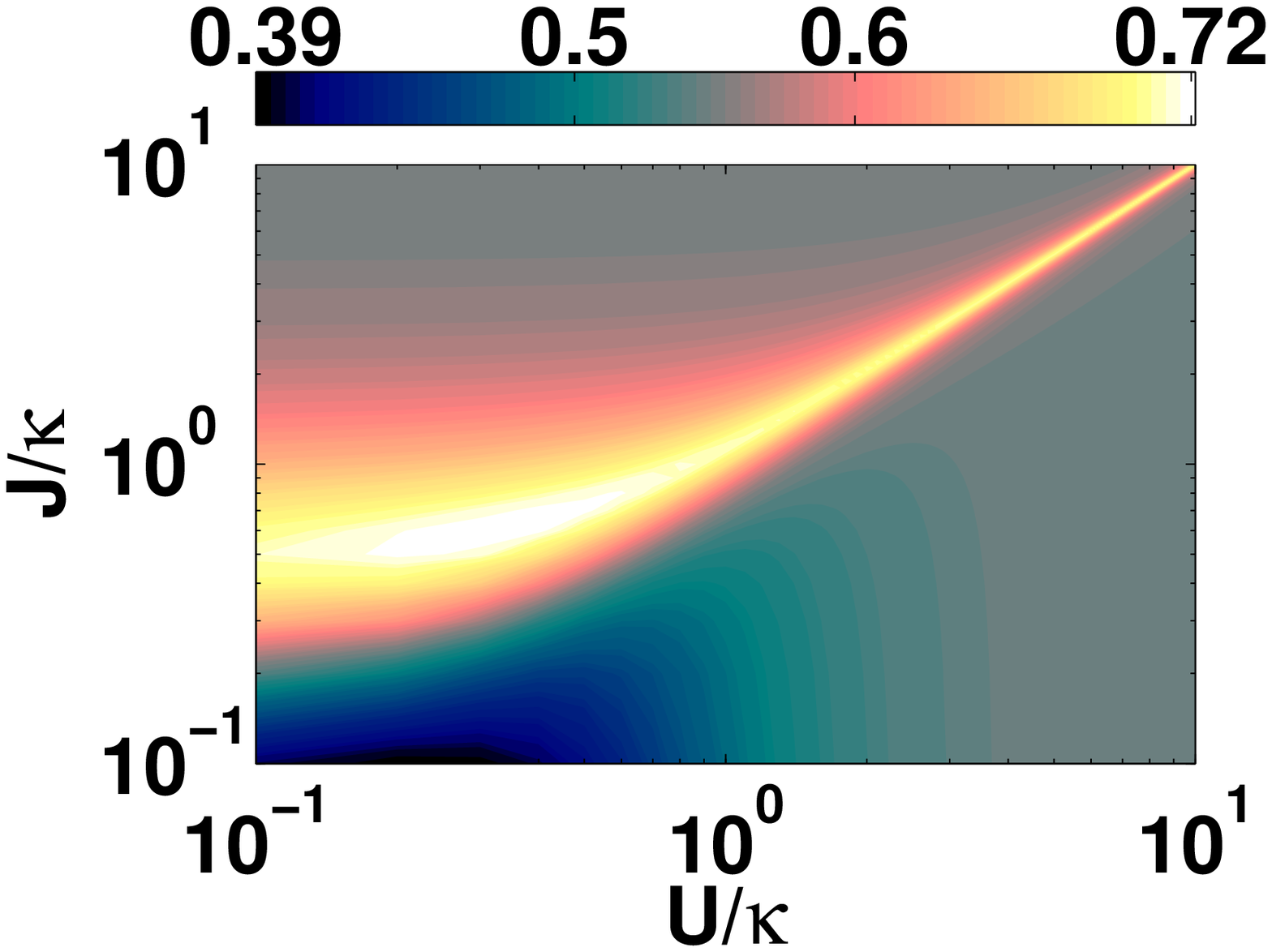}
        }%
        \subfigure[]{%
            \label{fig:fig4h}
            \includegraphics[width=4cm]{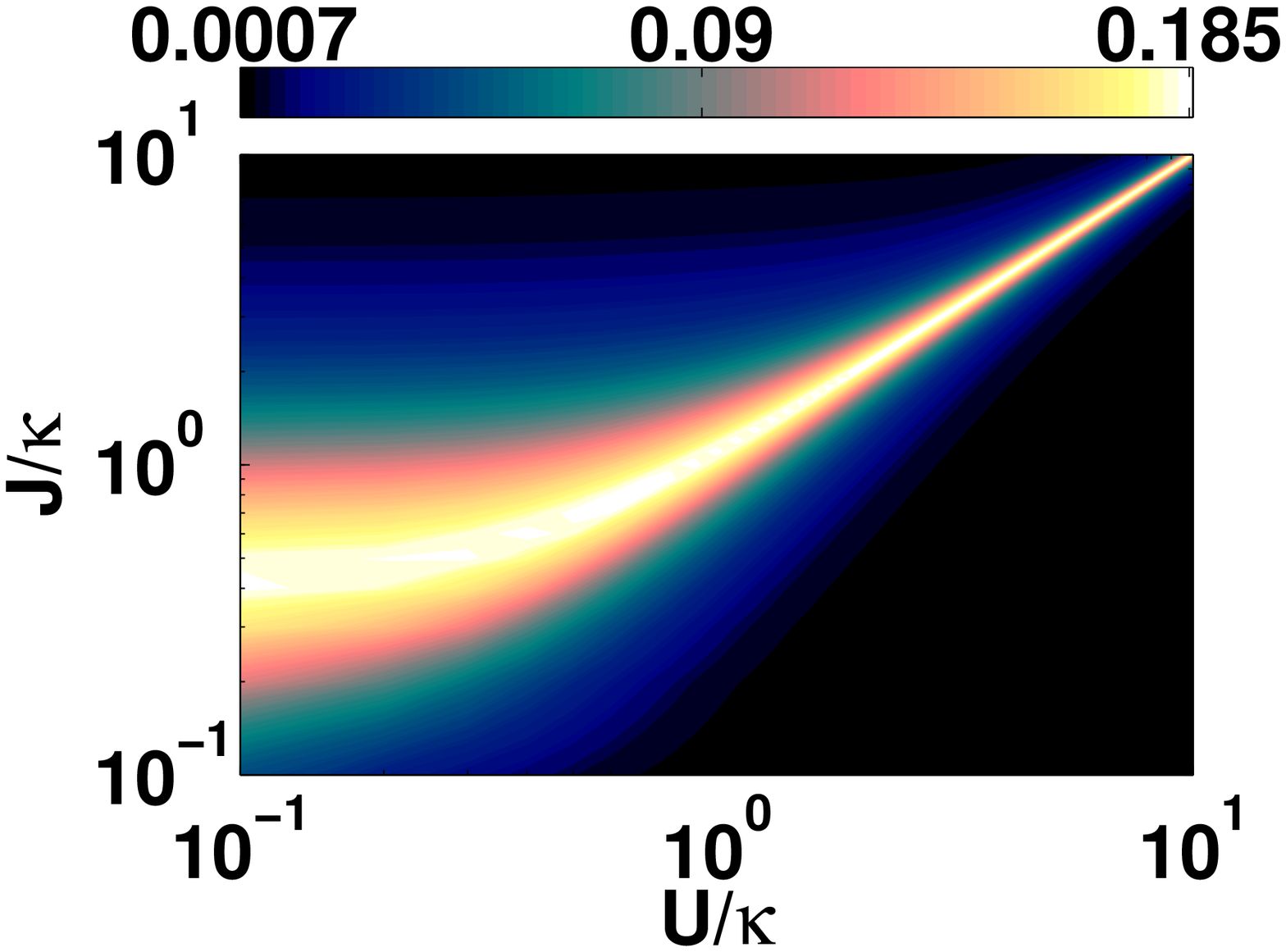}
        }%
    \end{center}
    \caption{%
        (Color Online) Dependence of (a) $\lambda_{1}$, (b) $\lambda_{2}$ (c) $S(\rho_{1})$ for $F/\kappa=0.1$ and (d) $\lambda_{1}$, (e) $\lambda_{2}$ (f) $S(\rho_{1})$ for $F/\kappa=1$ with respect to dimensionless $J/\kappa$ and $U/\kappa$ in two-site KH system with two-photon exchange. $\lambda_{1}$ and $\lambda_{2}$ are multiplied by $10^5$, $S(\rho_{1})$ is multiplied by $10^4$, and $E_N(\rho)$ is multiplied by $10^{3}$ in the case of $F/\kappa=0.1$ for visibility.
     }%
   \label{fig:fig4}
\end{figure*}
Mode entanglement parameter $\lambda_{1}$ behavior with $J$ and $U$, shown in Fig.~\ref{fig:fig2a}, is opposite to that of spin squeezing and concurrence. Localization of photons in the cavities favor pairwise particle entanglement; while wave like behavior associated with coherent delocalization by the photon exchange favors mode correlations. The behavior of the parameter $\lambda_{2}$, shown in the Fig.~\ref{fig:fig2b}, indicates that the mode entanglement occurs in the nonlocal regimes with strong nonlinearity. Fig.~\ref{fig:fig2e} and Fig.~\ref{fig:fig2f} show the mode entanglement parameters for the case of strong drive of $F/2\pi=0.4$ MHz. Here, mode correlations become harder to establish and regions where the modes are entangled are shifted towards higher $J/\kappa$ values.

Fig.~\ref{fig:fig2c} and Fig.~\ref{fig:fig2g} show the dependence of the von Neumann entropy for weak $F/2\pi=0.04$ MHz and strong $F/2\pi=0.4$ MHz drive conditions, respectively. Bipartite entanglement almost absent in the hopping dominant regimes, whereas the entropy reaches its maximal value in nonlinearity dominant regimes.

Fig.~\ref{fig:fig2d} and Fig.~\ref{fig:fig2h} show the behavior of logarithmic negativity $E_N(\rho)$. The impurity parameter for the weak drive case is calculated to be in the range $I\sim 0-10^{-4}$ while it becomes $I\sim 0-0.36$ for the strong drive case. We do not plot $I$ as it behaves as $g^{(2)}$. In the single photon exchange model, steady state is more pure in regions with greater Poissonian character. Only under strong drive and strong nonlinearity, where $I\sim 0.2-0.36$, characterization of the 
entanglement with $C_I$ and $S(\rho_1)$ can be ambiguous. In this case existence of entanglement is confirmed by $E_N$.

Behaviors of the genuine mode entanglement parameters $\lambda_{1}$ and $\lambda_{2}$ along with the von Neumann entropy give a complete description of mode correlations. First, occurrence of two-mode correlations in the sub-Poissonian regions is typical for two-squeezed light beams mixed with a beam splitter. Here, this happens simultaneously, not sequentially. Second, the von Neumann entropy covers the whole region which covered separately  by the mode-entanglement parameters. Bipartite entanglement is distributed over sub-Poissonian regimes.

We stressed out that the single photon exchange model gives clearly distinct roles to the interactions.
Nonlinearity is localizing and a single-axis twisting spin squeezing type of interaction. Photon hopping is delocalizing and a mode correlating interaction. Two photon hopping however has particle and delocalizing character. It is delocalizing due to photon exchange effect, but also establishes pairwise correlations by a two axis twisting interaction. We will point out emergence of distinct relation between second order coherence and entanglement in the following subsection.

\subsection{Two-Photon Exchange}\label{sec:two}

In Fig.~\ref{fig:fig3} and Fig.~\ref{fig:fig4}, we show our results for two-photon exchange interaction. Fig.~\ref{fig:fig3a}, Fig.~\ref{fig:fig3b} and Fig.~\ref{fig:fig3c} depict quantum coherence, spin squeezing and concurrence for $F/2\pi=0.04$ MHz, respectively.
In contrast to single photon exchange, there is no Poissonian or super-Poissonian region in Fig.~\ref{fig:fig3a} and $g^{(2)}(0)$ varies
in the range of $0.00723\leq g^{(2)}(0)\leq0.7944$. The sub-Poissonian statistics is strong in both delocalizing photon hopping and local cavity nonlinearity dominated regimes. Relatively weak one occurs in $J/\kappa,U/\kappa<1$ region where coherent drive is comparable to these local and nonlocal pairwise interactions.

In the two photon exchange case, analogous to LMG model, there are both single axis and two axis twisting routes to spin squeezing.
When $J/\kappa>1$ the former becomes the major route leading to spin squeezing; while when $U/\kappa>1$ the latter becomes the
main interaction causing spin squeezing. Accordingly strong violation of spin squeezing occurs over the entire domain of $J/\kappa,U/\kappa$. Similar level of violation can only be found in the narrow strip of weak $J/\kappa$ in Fig.~\ref{fig:fig1b}. The uniformity of the pairwise entanglement witness parameter with $U$ which is present in Fig.~\ref{fig:fig1b} is no longer present in Fig.~\ref{fig:fig3b}.
Variation of spin squeezing and sub-Poissonian statistics strength with hopping and nonlinearity becomes identical for the case of two photon exchange. As in the single-photon interaction case, concurrence is in complete agreement with spin squeezing. In both nonlinearity and hopping dominated regimes pairwise particle correlations show identical statistical character. Particle entanglement is maximized for $J/\kappa\gtrsim1$ and $U/\kappa\gtrsim1$.

Fig.~\ref{fig:fig3d}, Fig.~\ref{fig:fig3e} and Fig.~\ref{fig:fig3f} show quantum coherence, spin squeezing and concurrence for $F/2\pi=0.4$ MHz, respectively. Strong drive helps to bring coherence into the system and now the maximum $g^{(2)}(0)\sim 1$.
The separation of two strong sub-Poissonian regions by a weak one around
the line $J=U$ becomes more visible. Strong drive works against spin squeezing that reaches
its maximum value around $J/\kappa\sim0.5>U/\kappa$. Concurrence is maximized around similar $J/\kappa$ values
as spin squeezing with relatively strong nonlinearity.

In Fig.~\ref{fig:fig4}, we plot the mode entanglement and entropy parameters for the two-photon exchange interaction. Being only sufficient, values of the parameters $\lambda_{1}$ and $\lambda_{2}$ within our parameter regimes make the discussions on mode entanglement inconclusive~\cite{PhysRevA.87.022325}.

Fig.~\ref{fig:fig4c} shows the von Neumann entropy for the case of $F/2\pi=0.04$ MHz. Bipartite entanglement is strong in both local and nonlocal interaction regimes. In the case of strong pump, shown in Fig.~\ref{fig:fig4f}, it is maximized in the diagonal which separates two weaker regimes. This diagonal corresponds to the weaker multiparticle but stronger pairwise entanglement situations depicted in Fig.~\ref{fig:fig3e} and Fig.~\ref{fig:fig3f}, respectively.

Fig.~\ref{fig:fig4d} and Fig.~\ref{fig:fig4h} show the behavior of logarithmic negativity $E_N(\rho)$. The impurity parameter for the weak drive case is calculated to be in the range $I\sim 10^{-5}-10^{-4}$ while it becomes $I\sim 0.22-0.39$ for the strong drive case. We do not plot $I$ as it behaves as $g^{(2)}$. In the two-photon exchange model, similar to the single photon exchange case, steady state is more pure in regions with greater Poissonian character. In contrast to single photon exchange case however, mixed state region with
$I\sim 0.2-0.36$ is not limited to strong nonlinearity but extends 
over the entire region under strong drive and hence characterization of the 
entanglement with $C_I$ and $S(\rho_1)$ can be ambiguous. In this case existence of entanglement is confirmed by $E_N$.
\section{Conclusion}\label{sec:conc}
We finally present a summary of our comparative analysis and associated conclusions. 
We considered single photon and two-photon exchange coupled cavities with Kerr nonlinearity under drive and dissipation. We examined the steady state second order coherence, multi-particle and pairwise entanglement as well as bipartite and mode entanglement parameters.

Two-photon exchange between weakly driven identical nonlinear cavities make the photons strongly antibunched for any $J$ and $U$, except for weak $J$ and $U$ quadrant where photons have weaker sub-Poissonian character. Single photon exchange can only generate antibunched photons for $J<U$ and coherent photons for $J>U$~\cite{ferretti2010photon}. A strong drive is most influential to enhance the coherence regime for two photon hopping case when $J\sim U$.

Our analysis reveals viable routes to entanglement realization in our models. Coherence serves as a map to search for most advantageous J and U domains for practical establishment of entanglement. Under weak drive, antibunched photons in single photon exchange violate the product state conditions $3-5$ orders of magnitude more strongly than the super-Poisson ones. Two-photon exchange yields almost uniform violation with similar but maximum quantum noise levels. This points out one crucial advantage of two photon exchange over the single photon exchange for practical realization of entanglement. There are both single and two-axis twisting spin squeezing routes to entanglement in two-photon exchange which leads to almost uniform and strong satisfaction of particle, pairwise and mode entanglement conditions over the entire $J, U$ domain. There is another surprising and appealing advantage in the two-photon exchange under strong drive. Entanglement emerges most easily in weak nonlinearity and weak hopping region and with photons of relatively weak sub-Poisson character. 

For both single and two-photon exchange, multi-particle entanglement is harder to establish relative to pairwise and mode entanglements, which are at similar levels of quantum noise. Quantum fluctuations could be too low to detect in experiments for all types of entanglements under weak drive. Strong drive remedy this situation and increases quantum fluctuations $1-2$ orders of magnitude more in steady state. Strong drive also leads to mixed state entanglement, albeit with weak mixedness, in the sub-Poisson regime of single photon exchange or everywhere in two-photon exchange.

Our results illuminates the subtle relations between locality, coherence and quantum correlations that could be exploited for synthesis of many body quantum entangled states in coupled cavities with local and nonlocal nonlinearities.
\acknowledgements
We thank P.~Forn-D\'{\i}az and J.~Vidal for illuminating discussions.
\"O. E. M. acknowledges financial support from the National Science and Technology Foundation of Turkey (T\"UBITAK)
(Grant No. 111T285).

\begin{thebibliography}{10}
\newcommand{\enquote}[1]{``#1''}

\bibitem{hillery2009introduction}
M.~Hillery, \enquote{An introduction to the quantum theory of nonlinear
  optics,} Acta Phys. Slovaca \textbf{59}, 1--80 (2009).

\bibitem{dell2006multiphoton}
F.~Dell{\'\i}Anno, S.~De~Siena, and F.~Illuminati, \enquote{Multiphoton quantum
  optics and quantum state engineering,} Phys. Rep. \textbf{428}, 53--168
  (2006).

\bibitem{ham2000coherence}
B.~S. Ham and P.~R. Hemmer, \enquote{Coherence switching in a four-level
  system: quantum switching,} Phys. Rev. Lett. \textbf{84}, 4080--4083 (2000).

\bibitem{nielsen2010quantum}
M.~A. Nielsen and I.~L. Chuang, \emph{Quantum Computation and Quantum
  Information} (Cambridge university press, 2010).

\bibitem{vidal2003entanglement}
G.~Vidal, J.~I. Latorre, E.~Rico, and A.~Kitaev, \enquote{Entanglement in
  quantum critical phenomena,} Phys. Rev. Lett. \textbf{90}, 227902 (2003).

\bibitem{sondhi1997continuous}
S.~Sondhi, S.~Girvin, J.~Carini, and D.~Shahar, \enquote{Continuous quantum
  phase transitions,} Rev. Mod. Phys. \textbf{69}, 315 (1997).

\bibitem{PhysRevLett.109.053601}
M.~Schir\'o, M.~Bordyuh, B.~\"Oztop, and H.~E. T\"ureci, \enquote{Phase
  transition of light in cavity qed lattices,} Phys. Rev. Lett. \textbf{109},
  053601 (2012).

\bibitem{PhysRevA.76.031805}
D.~G. Angelakis, M.~F. Santos, and S.~Bose, \enquote{Photon-blockade-induced
  mott transitions and $xy$ spin models in coupled cavity arrays,} Phys. Rev. A
  \textbf{76}, 031805 (2007).

\bibitem{PhysRevLett.99.186401}
D.~Rossini and R.~Fazio, \enquote{Mott-insulating and glassy phases of
  polaritons in 1d arrays of coupled cavities,} Phys. Rev. Lett. \textbf{99},
  186401 (2007).

\bibitem{PhysRevLett.100.216401}
M.~Aichhorn, M.~Hohenadler, C.~Tahan, and P.~B. Littlewood, \enquote{Quantum
  fluctuations, temperature, and detuning effects in solid-light systems,}
  Phys. Rev. Lett. \textbf{100}, 216401 (2008).

\bibitem{PhysRevA.77.031803}
N.~Na, S.~Utsunomiya, L.~Tian, and Y.~Yamamoto, \enquote{Strongly correlated
  polaritons in a two-dimensional array of photonic crystal microcavities,}
  Phys. Rev. A \textbf{77}, 031803 (2008).

\bibitem{PhysRevLett.103.086403}
S.~Schmidt and G.~Blatter, \enquote{Strong coupling theory for the
  jaynes-cummings-hubbard model,} Phys. Rev. Lett. \textbf{103}, 086403 (2009).

\bibitem{PhysRevA.80.023811}
J.~Koch and K.~Le~Hur, \enquote{Superfluid\char21{}mott-insulator transition of
  light in the jaynes-cummings lattice,} Phys. Rev. A \textbf{80}, 023811
  (2009).

\bibitem{PhysRevA.80.033612}
P.~Pippan, H.~G. Evertz, and M.~Hohenadler, \enquote{Excitation spectra of
  strongly correlated lattice bosons and polaritons,} Phys. Rev. A \textbf{80},
  033612 (2009).

\bibitem{PhysRevA.81.061801}
A.~Tomadin, V.~Giovannetti, R.~Fazio, D.~Gerace, I.~Carusotto, H.~E. T\"ureci,
  and A.~Imamoglu, \enquote{Signatures of the superfluid-insulator phase
  transition in laser-driven dissipative nonlinear cavity arrays,} Phys. Rev. A
  \textbf{81}, 061801 (2010).

\bibitem{PhysRevLett.104.216402}
S.~Schmidt and G.~Blatter, \enquote{Excitations of strongly correlated lattice
  polaritons,} Phys. Rev. Lett. \textbf{104}, 216402 (2010).

\bibitem{hartmann2006strongly}
M.~J. Hartmann, F.~G. Brand{\~a}o, and M.~B. Plenio, \enquote{Strongly
  interacting polaritons in coupled arrays of cavities,} Nature Physics
  \textbf{2}, 849--855 (2006).

\bibitem{greentree2006quantum}
A.~D. Greentree, C.~Tahan, J.~H. Cole, and L.~C. Hollenberg, \enquote{Quantum
  phase transitions of light,} Nature Physics \textbf{2}, 856--861 (2006).

\bibitem{PhysRevLett.93.037001}
S.~Saito, M.~Thorwart, H.~Tanaka, M.~Ueda, H.~Nakano, K.~Semba, and
  H.~Takayanagi, \enquote{Multiphoton transitions in a macroscopic quantum
  two-state system,} Phys. Rev. Lett. \textbf{93}, 037001 (2004).

\bibitem{ferretti2010photon}
S.~Ferretti, L.~C. Andreani, H.~E. T{\"u}reci, and D.~Gerace, \enquote{Photon
  correlations in a two-site nonlinear cavity system under coherent drive and
  dissipation,} Phys. Rev. A \textbf{82}, 013841 (2010).

\bibitem{paul1982photon}
H.~Paul, \enquote{Photon antibunching,} Rev. Mod. Phys. \textbf{54}, 1061
  (1982).

\bibitem{horodecki2009quantum}
R.~Horodecki, P.~Horodecki, M.~Horodecki, and K.~Horodecki, \enquote{Quantum
  entanglement,} Rev. Mod. Phys. \textbf{81}, 865 (2009).

\bibitem{alexanian2011two}
M.~Alexanian, \enquote{Two-photon exchange between two three-level atoms in
  separate cavities,} Physical Review A \textbf{83}, 023814 (2011).

\bibitem{Hardal12}
A.~{\"U}.~C. Hardal and {\"O}.~E. M{\"u}stecapl{\i}o{\u{g}}lu,
  \enquote{Transfer of spin squeezing and particle entanglement between atoms
  and photons in coupled cavities via two-photon exchange,} J. Opt. Soc. Am. B
  \textbf{29}, 1822--1828 (2012).

\bibitem{PhysRevA.85.023833}
Y.-L. Dong, S.-Q. Zhu, and W.-L. You, \enquote{Quantum-state transmission in a
  cavity array via two-photon exchange,} Phys. Rev. A \textbf{85}, 023833
  (2012).

\bibitem{PhysRevLett.105.100505}
L.~S. Bishop, E.~Ginossar, and S.~M. Girvin, \enquote{Response of the strongly
  driven jaynes-cummings oscillator,} Phys. Rev. Lett. \textbf{105}, 100505
  (2010).

\bibitem{PhysRevA.46.R6801}
L.~Tian and H.~J. Carmichael, \enquote{Quantum trajectory simulations of
  two-state behavior in an optical cavity containing one atom,} Phys. Rev. A
  \textbf{46}, R6801--R6804 (1992).

\bibitem{alexanian2010scattering}
M.~Alexanian, \enquote{Scattering of two coherent photons inside a
  one-dimensional coupled-resonator waveguide,} Phys. Rev. A \textbf{81},
  015805 (2010).

\bibitem{PhysRevB.79.180511}
P.~J. Leek, S.~Filipp, P.~Maurer, M.~Baur, R.~Bianchetti, J.~M. Fink,
  M.~G\"oppl, L.~Steffen, and A.~Wallraff, \enquote{Using sideband transitions
  for two-qubit operations in superconducting circuits,} Phys. Rev. B
  \textbf{79}, 180511 (2009).

\bibitem{PhysRevB.77.180502}
J.~A. Schreier, A.~A. Houck, J.~Koch, D.~I. Schuster, B.~R. Johnson, J.~M.
  Chow, J.~M. Gambetta, J.~Majer, L.~Frunzio, M.~H. Devoret, S.~M. Girvin, and
  R.~J. Schoelkopf, \enquote{Suppressing charge noise decoherence in
  superconducting charge qubits,} Phys. Rev. B \textbf{77}, 180502 (2008).

\bibitem{niemczyk2011selection}
T.~Niemczyk, F.~Deppe, E.~Menzel, M.~Schwarz, H.~Huebl, F.~Hocke,
  M.~H{\"a}berlein, M.~Danner, E.~Hoffmann, A.~Baust \emph{et~al.},
  \enquote{Selection rules in a strongly coupled qubit-resonator system,}
  arXiv:1107.0810  (2011).

\bibitem{deppe2008two}
F.~Deppe, M.~Mariantoni, E.~Menzel, A.~Marx, S.~Saito, K.~Kakuyanagi,
  H.~Tanaka, T.~Meno, K.~Semba, H.~Takayanagi \emph{et~al.},
  \enquote{Two-photon probe of the jaynes--cummings model and controlled
  symmetry breaking in circuit qed,} Nature Physics \textbf{4}, 686--691
  (2008).

\bibitem{mariantoni2011photon}
M.~Mariantoni, H.~Wang, R.~C. Bialczak, M.~Lenander, E.~Lucero, M.~Neeley,
  A.~O{\'\i}Connell, D.~Sank, M.~Weides, J.~Wenner \emph{et~al.},
  \enquote{Photon shell game in three-resonator circuit quantum
  electrodynamics,} Nature Physics \textbf{7}, 287--293 (2011).

\bibitem{ma2011quantum}
J.~Ma, X.~Wang, C.~Sun, and F.~Nori, \enquote{Quantum spin squeezing,} Phys.
  Rep.  (2011).

\bibitem{marchiolli2013spin}
M.~A. Marchiolli, D.~Galetti, and T.~Debarba, \enquote{Spin squeezing and
  entanglement via finite-dimensional discrete phase-space description,}
  International Journal of Quantum Information \textbf{11}, 1330001 (2013).

\bibitem{wang_spin_2003}
X.~Wang and B.~C. Sanders, \enquote{Spin squeezing and pairwise entanglement
  for symmetric multiqubit states,} Physical Review A \textbf{68}, 012101
  (2003).

\bibitem{dong_spin_2005}
Y.~Dong, W.~Xiao-Guang, and W.~Ling-An, \enquote{Spin squeezing and
  entanglement of many-particle spin-half states,} Chinese Physics Letters
  \textbf{22}, 271 (2005).

\bibitem{mustecap2011}
{\"O}.~E. M{\"u}stecapl{\i}o{\u{g}}lu, \enquote{Quantum coherence and
  correlations of optical radiation by atomic ensembles interacting with a
  two-level atom in a microwave cavity,} Phys. Rev. A \textbf{83}, 023805
  (2011).

\bibitem{lipkin1965validity}
H.~Lipkin, N.~Meshkov, and A.~Glick, \enquote{Validity of many-body
  approximation methods for a solvable model::(i). exact solutions and
  perturbation theory,} Nuclear Physics \textbf{62}, 188--198 (1965).

\bibitem{PhysRevA.70.062304}
J.~Vidal, G.~Palacios, and C.~Aslangul, \enquote{Entanglement dynamics in the 
Lipkin-Meshkov-Glick model,} Phys. Rev. A \textbf{70}, 062304 (2004).

\bibitem{PhysRevE.78.021106}
P.~Ribeiro, J.~Vidal, and R.~Mosseri, \enquote{Exact spectrum of the 
Lipkin-Meshkov-Glick model in the thermodynamic limit and finite-size
  corrections,} Phys. Rev. E \textbf{78}, 021106 (2008).

\bibitem{PhysRevLett.99.050402}
P.~Ribeiro, J.~Vidal, and R.~Mosseri, \enquote{Thermodynamical limit of the 
Lipkin-Meshkov-Glick model,} Phys. Rev. Lett. \textbf{99}, 050402 (2007).

\bibitem{scully1997quantum}
M.~O. Scully and M.~S. Zubairy, \emph{Quantum Optics} (Cambridge University
  Press, Cambridge, 1997).

\bibitem{PhysRevLett.99.250405}
G.~T\'oth, C.~Knapp, O.~G\"uhne, and H.~J. Briegel, \enquote{Optimal spin
  squeezing inequalities detect bound entanglement in spin models,} Phys. Rev.
  Lett. \textbf{99}, 250405 (2007).

\bibitem{PhysRevA.64.042315}
P.~Rungta, V.~Bu\ifmmode~\check{z}\else \v{z}\fi{}ek, C.~M. Caves, M.~Hillery,
  and G.~J. Milburn, \enquote{Universal state inversion and concurrence in
  arbitrary dimensions,} Phys. Rev. A \textbf{64}, 042315 (2001).

\bibitem{PhysRevLett.96.050503}
M.~Hillery and M.~S. Zubairy, \enquote{Entanglement conditions for two-mode
  states,} Phys. Rev. Lett. \textbf{96}, 050503 (2006).

\bibitem{PhysRevA.74.032333}
M.~Hillery and M.~S. Zubairy, \enquote{Entanglement conditions for two-mode
  states: Applications,} Phys. Rev. A \textbf{74}, 032333 (2006).

\bibitem{cunha2007entanglement}
M.~O.~T. Cunha, J.~A. Dunningham, and V.~Vedral, \enquote{Entanglement in
  single-particle systems,} Proceedings of the Royal Society A: Mathematical,
  Physical and Engineering Science \textbf{463}, 2277--2286 (2007).

\bibitem{PhysRevA.72.064306}
S.~J. van Enk, \enquote{Single-particle entanglement,} Phys. Rev. A
  \textbf{72}, 064306 (2005).

\bibitem{benatti2011entanglement}
F.~Benatti, R.~Floreanini, and U.~Marzolino, \enquote{Entanglement and
  squeezing with identical particles: ultracold atom quantum metrology,}
  Journal of Physics B: Atomic, Molecular and Optical Physics \textbf{44},
  091001 (2011).

\bibitem{prevedel2009experimental}
R.~Prevedel, G.~Cronenberg, M.~Tame, M.~Paternostro, P.~Walther, M.~Kim, and
  A.~Zeilinger, \enquote{Experimental realization of dicke states of up to six
  qubits for multiparty quantum networking,} Phys. Rev. Lett. \textbf{103},
  20503 (2009).

\bibitem{barreiro2010experimental}
J.~T. Barreiro, P.~Schindler, O.~G{\"u}hne, T.~Monz, M.~Chwalla, C.~F. Roos,
  M.~Hennrich, and R.~Blatt, \enquote{Experimental multiparticle entanglement
  dynamics induced by decoherence,} Nature Physics \textbf{6}, 943--946 (2010).

\bibitem{korbicz2006generalized}
J.~Korbicz, O.~G{\"u}hne, M.~Lewenstein, H.~H{\"a}ffner, C.~Roos, and R.~Blatt,
  \enquote{Generalized spin-squeezing inequalities in n-qubit systems: Theory
  and experiment,} Phys. Rev. A \textbf{74}, 052319 (2006).

\bibitem{hartmann08}
M.~J.~Hartmann, F.~G.~S.~L.~Brand\~{a}o, and M.~B.~Plenio,
\endquote{Quantum many-body phenomena in coupled cavity arrays,}
Laser Photon.~Rev. \textbf{2}, 527-556 (2008).

\bibitem{liew12}
T.~C.~H.~Liew and V.~Savona,
\endquote{Quantum entanglement in nanocavity arrays,}
Phys.~Rev.~A \textbf{85}, 050301 (2012).

\bibitem{zhou2009pair}
X.~Zhou, Y.~Zhang, and G.~Guo, \enquote{Pair tunneling of bosonic atoms in an
  optical lattice,} Physical Review A \textbf{80}, 013605 (2009).

\bibitem{PhysRevLett.81.1539}
D.~S. Hall, M.~R. Matthews, J.~R. Ensher, C.~E. Wieman, and E.~A. Cornell,
  \enquote{Dynamics of component separation in a binary mixture of
  bose-einstein condensates,} Phys. Rev. Lett. \textbf{81}, 1539--1542 (1998).

\bibitem{PhysRevLett.81.1543}
D.~S. Hall, M.~R. Matthews, C.~E. Wieman, and E.~A. Cornell,
  \enquote{Measurements of relative phase in two-component bose-einstein
  condensates,} Phys. Rev. Lett. \textbf{81}, 1543--1546 (1998).

\bibitem{PhysRevLett.98.030402}
B.~V. Hall, S.~Whitlock, R.~Anderson, P.~Hannaford, and A.~I. Sidorov,
  \enquote{Condensate splitting in an asymmetric double well for atom chip
  based sensors,} Phys. Rev. Lett. \textbf{98}, 030402 (2007).

\bibitem{ketterle_nature}
J.~Stenger, S.~Inouye, D.~Stamper-Kurn, H.~Miesner, A.~Chikkatur, and
  W.~Ketterle, \enquote{Spin domains in ground-state bose-einstein
  condensates,} NATURE \textbf{396}, 345--348 (1998).

\bibitem{Pietraszewicz12}
J.~Pietraszewicz, T.~Sowi\'{n}ski, M.~Brewczyk, J.~Zakrzewski, M.~Lewenstein, and M.~Gajda,\endquote{Two-component Bose-Hubbard model with higher-angular-momentum states}
Phys. Rev. A
\textbf{85}, 053638 (2012).

\bibitem{greentree06}
A.~D.~Greentree, C.~Tahan, J.~H.~Cole, and L.~C.~L.~Hollenberg,
\endquote{Quantum phase transitions of light,}
Nat.~Phys. \textbf{2}, 856-861 (2006).

\bibitem{jin13}
J.~Jin, D.~Rossini, R.~Fazio, M.~Leib, and M.~Hartmann,\endquote{Photon solid phases in driven arrays of nonlinearly coupled cavities,}
Phys.~Rev.~Lett. \textbf{110}, 163605 (2013).

\bibitem{PhysRevA.47.5138}
M.~Kitagawa and M.~Ueda, \enquote{Squeezed spin states,} Phys. Rev. A
  \textbf{47}, 5138--5143 (1993).

\bibitem{PhysRevA.79.013819}
M.~Boissonneault, J.~M.~Gambetta, and A.~Blais,
\enquote{Dispersive regime of circuit QED: Photon-dependent qubit dephasing and relaxation rates,} 
Phys. Rev. A
\textbf{79}, 
013819 
(2009).

\bibitem{johansson_qutip_2013}
J.~Johansson, P.~Nation, and F.~Nori, \enquote{{QuTiP} 2: A python framework
  for the dynamics of open quantum systems,} Computer Physics Communications
  \textbf{184}, 1234--1240 (2013).

\bibitem{PhysRevLett.105.140502}
Y.~Kubo, F.~R. Ong, P.~Bertet, D.~Vion, V.~Jacques, D.~Zheng, A.~Dr\'eau, J.-F.
  Roch, A.~Auffeves, F.~Jelezko, J.~Wrachtrup, M.~F. Barthe, P.~Bergonzo, and
  D.~Esteve, \enquote{Strong coupling of a spin ensemble to a superconducting
  resonator,} Phys. Rev. Lett. \textbf{105}, 140502 (2010).

\bibitem{PhysRevA.75.032329}
A.~Blais, J.~Gambetta, A.~Wallraff, D.~I. Schuster, S.~M. Girvin, M.~H.
  Devoret, and R.~J. Schoelkopf, \enquote{Quantum-information processing with
  circuit quantum electrodynamics,} Phys. Rev. A \textbf{75}, 032329 (2007).

\bibitem{vidal2006concurrence}
J.~Vidal, \enquote{Concurrence in collective models,} Physical Review A
  \textbf{73}, 062318 (2006).

\bibitem{PhysRevA.87.022325}
B.~Sen, S.~K. Giri, S.~Mandal, C.~H.~R. Ooi, and A.~Pathak, \enquote{Intermodal
  entanglement in raman processes,} Phys. Rev. A \textbf{87}, 022325 (2013).
  
\bibitem{vidal02}
G.~Vidal and R.~F.~Werner,
\enquote{Computable measure of entanglement,} 
Phys. Rev. A \textbf{65}, 032314 (2002).
  
\bibitem{plenio05}
M.~B.~Plenio, 
\enquote{Logarithmic negativity: a full entanglement monotone that is not convex,} 
Phys. Rev. Lett. \textbf{95}, 090503 (2005).

\end{thebibliography}

\end{document}